\documentclass[aps,twocolumn,prb, nofootinbib]{revtex4-1}
\usepackage[normalem]{ulem}
\usepackage{amsmath,amstext,amssymb, graphicx,bm,dcolumn,color,times, braket,inputenc}
\usepackage[colorlinks=true,citecolor=blue,urlcolor=blue]{hyperref}
\usepackage[table]{xcolor}

\usepackage{natbib}

\begin{document}
\title{Phase boundaries in alternating field quantum XY model with Dzyaloshinskii-Moriya interaction: Sustainable entanglement in dynamics}

\author{Saptarshi Roy${}^{1}$, Titas Chanda${}^{1,2}$, Tamoghna Das${}^{1, 3}$, Debasis Sadhukhan${}^{1,2}$, Aditi Sen(De)${}^{1}$, Ujjwal Sen${}^{1}$}

\affiliation{${}^{1}$Harish-Chandra Research Institute, HBNI, Chhatnag Road, Jhunsi, Allahabad 211 019, India \\
${}^{2}$ Instytut Fizyki im. Mariana Smoluchowskiego, Uniwersytet Jagiello\'nski,  \L{}ojasiewicza 11, 30-348 Krak\'ow, Poland,\\ 
${}^3$Institute of Informatics, National Quantum Information Centre, Faculty of Mathematics, Physics and Informatics, 
University of Gdan\'{s}k, 80-308 Gdan\'{s}k, Poland
}

\begin{abstract}
We report all phases and corresponding critical lines of the quantum anisotropic transverse XY model with uniform and alternating transverse magnetic fields (ATXY) in presence of the Dzyaloshinskii-Moriya (DM) interaction by using appropriately chosen order parameters. We prove that when DM interaction is weaker than the anisotropy parameter, it has no effect at all on the zero-temperature states of the XY model with uniform transverse magnetic field (UXY), which is not the case for the ATXY model. However, when DM interaction is stronger than the anisotropy parameter, we show appearance of a new \emph{gapless} chiral phase - in the XY model with uniform as well as alternating field. We further observe that first derivatives of 
nearest neighbor two-site entanglement with respect to magnetic fields can detect all the critical lines present in the system. 
We also find that the factorization surface at zero-temperature present in this model without DM interaction becomes a volume on the introduction of the later.
Moreover, DM interaction turns out to be good to generate  bipartite entanglement sustainable at large times, leading to a proof of ergodic nature of bipartite entanglement in this system.

\end{abstract}

\maketitle

\section{Introduction}
Quantum phase transitions \cite{qptbook1, qptbook2, qptbook3}, observed at zero temperature in  many-body systems, are one of the striking phenomena  in quantum mechanics which occurs solely due to quantum fluctuations in the system. 
Detecting such transitions in spin systems have attracted lot of attentions over decades.  Machinery borrowed from quantum information
theory  \cite{dmrg1, dmrg2, dmrg3, dmrg_ex1, dmrg_ex2, dmrg_ex3} have proven to be useful in developing new
techniques to obtain zero-temperature state in interacting spin
Hamiltonian.
On the experimental front,  tremendous scientific and technological advancement in cold atomic systems \cite{cold_atom1, cold_atom2, cold_atom3, cold_atom4, cold_atom5, cold_atom6, cold_atom7, sci_exp}, superconducting materials \cite{sup_cond}, nuclear magnetic resonance (NMR) molecules \cite{nmr1}  (see also \cite{nmr2, nmr3, nmr4, nmr5, nmr6}) open up the possibilities to realize   and examine properties of such many-body systems in the laboratory (see also \cite{nat_exp}).

On the other hand, it has been established over the last three decades that quantum entanglement \cite{H4} is  an essential ingredient in  quantum information processing tasks \cite{DC, TP, QKD,FP1, FP2, FP3} which are more efficient than their classical counterparts.  
Highly entangled ground states of interacting spin systems, which are realizable in currently available technology \cite{ADP_Aditi, AmicoRMP, Area}, turn out to be natural candidates for realizing such quantum protocols.
Moreover, besides the conventional order parameters, quantum entanglement has emerged as an independent tool to identify the signature of quantum criticality in quantum spin models \cite{AmicoRMP, ADP_Aditi, Area}.
Apart from the fundamental importance of studying zero-temperature and thermal equilibrium states,  dynamical quantum correlations, generated in  the many-body systems in non-equilibrium scenarios, have
also been proven to be important in different directions, like topological quantum computation \cite{topo_comp1, topo_comp2, topo_comp3, topo_comp4, topo_comp5, topo_comp6, topo_comp7, topo_comp8}, observation of dynamical phase transitions \cite{dyncorr1, dyncorr2, dyncorr3, dyncorr4, dyncorr5, dyncorr6, dyncorr7, dyncorr8, dyncorr9, dyncorr10, dyncorr11, dyncorr12, dyncorr13, dyncorr14, dyncorr15, dyncorr16, dyncorr17}, answering statistical-mechanical questions like ergodicity of  quantum observables \cite{bm1, bm2, ergo_klein, ergo_loinger, atxy_pra, amaderkaj1, dyncorr2, amaderkaj3, amaderkaj4, amaderkaj5} etc.

However, most of the studies in this direction are restricted to models with symmetric spin-spin interactions such as Ising, XY, Heisenberg etc. \cite{ADP_Aditi, AmicoRMP}.
But works of Dzyaloshinskii \cite{dm_d}, Moriya \cite{dm_moriya1, dm_moriya2} and Anderson \cite{dm_anderson}, prompt one to consider asymmetric spin-spin interactions, e.g. Dzyaloshinskii-Moriya (DM) interaction, to explain the presence of weak ferromagnetism  in certain materials like  $\alpha-\text{Fe}_2\text{O}_3$, $\text{MnCO}_3$ which are bulk antiferromagnets. In a general context, DM interaction leads to novel phases which break the mirror inversion symmetry (chirality), and thus have paved the way for lots of research \cite{jetp, dm1, dm2, dm3, dm4, dm5, dm6, dm7,dm8, dm9, dm10, dm11, dm12, dm13, dm14, dm15, dm16, dm17, dm18, dm19, dm20, dm21, dm22, dm23, dm24, dm25, dm26, dm27, dm28, dm29, dm30, dm31, pre_duxy}. Like some other one-dimensional quantum spin models, spin chain with 
DM interaction can in certain instances be mapped to a Hamiltonian of spinless fermions \cite{JW} or hardcore bosons \cite{HCB-HP} and thus can also be realized e.g in cold atoms
\cite{bec_expt, dm_fazio, dm24, monica} as well as in nuclear magnetic resonance (NMR) systems \cite{anilkumar1, anilkumar2}, thereby providing possibilities to probe the effects of DM interaction  in laboratories.

In this paper, we investigate the effects of DM interaction on the quantum XY spin model with uniform and alternating transverse magnetic fields (ATXY) in one dimension. We identify quantum critical lines by gap closing in the energy spectrum as well as by the first derivatives of entanglement and the corresponding phases by using appropriate order parameters. 
Specifically, when DM interaction is weaker than the anisotropy parameter of the exchange coupling, we establish three distinct phases  -- antiferromagnetic  (AFM), paramagnetic-I (PM-I) and paramagnetic-II (PM-II) in which only AFM to PM-II transition depends on the strength of the DM interaction. 
In the case of quantum XY model only with uniform transverse field (UXY), 
we prove that,  when the strength of the DM interaction is strictly less than the anisotropy parameter, the system at the zero-temperature is insensitive to the  DM interaction, and hence all the physical properties including entanglement of the zero-temperature state remain unaltered. 
However, this is not the case when the alternating field
is introduced or when the DM interaction is stronger than the
anisotropy parameter in the $x-y$ direction.
Although DM interactions were typically found to be weak compared to other nearest-neighbor interactions, recent theoretical as well as experimental investigations show that the ratio between DM  and  other interactions can be finite \cite{dmf1, dmf2, dmf3, dmf4}. 
Motivated by these results, we  consider finite DM interaction  in the ATXY model, and report a new phase, the gapless chiral (CH) phase, that  emerges in place of the AFM one.  We also find two quantum critical lines --  CH to PM-I and CH to PM-II, both of which depend on the strength of the DM interaction. 

We also observe that bipartite entanglement and its first derivatives with respect to system parameters  can faithfully detect all the phase-boundaries. Moreover, we find that the lower value of entanglement  in the AFM phase can be enhanced by increasing the strength of the DM interaction which is possibly due to the appearance of the chiral phase. We also find that with the introduction of weak as well as strong DM interaction, the  factorization surface, i.e., the surface where entanglement vanishes in the zero-temperature state of the ATXY model, becomes a volume.   In case of the thermal state, DM interaction induces a transition from monotonic variation of entanglement with temperature to a nonmonotonic one and vice-versa. 

In the dynamical evolution of the system after a sudden quench, a high value of bipartite entanglement is found to be generated at a small time which ultimately saturates to a positive value at large time. Interestingly, we observe that the presence of the DM interaction enhances the saturation value in dynamics, thereby establishing its capability in realizations of quantum information tasks.  
From a statistical mechanical point of view, 
we show here that moderate DM interaction wipes out the nonergodic nature of bipartite entanglement, leading to ergodicity of entanglement irrespective of quantum phases (cf. \cite{atxy_pra}).

The paper is organized as follows: Sec. \ref{sec_model} discusses the diagonalization procedure of the Hamiltonian considered. In Subsec. \ref{uniformfield}, we discuss all the analytical calculations for the uniform field case. Sec. \ref{sec_phases} contains the characterization of different phases for different DM interaction strengths by using suitable order parameters. 
The detection of phase boundaries by entanglement, the effects of DM interaction in the thermal state, the calculation for the factorization volume in this model are reported in Sec. \ref{sec:DMent} We discuss the sudden quenched dynamics of entanglement and consequently its ergodicity property  in Sec. \ref{sec_dyn}, before the concluding remarks in  Sec. \ref{sec_discuss}. 

\section{Model and its diagonalization}
\label{sec_model} 
We consider a family of interacting spin models consisting of spin-$1/2$ particles on a one dimensional (1D) lattice with $N$ sites, described by a Hamiltonian, 
\small
\begin{eqnarray}
&\hat{H}&= \frac{1}{2}\sum_{j = 1}^{N}\Big[J\Big\{\frac{1+\gamma}{2}\hat{\sigma}_j^x\hat{\sigma}_{j+1}^{x}+\frac{1-\gamma}{2}\hat{\sigma}_j^y\hat{\sigma}_{j+1}^{y}\Big\} \nonumber \\ 
&+& \frac{D}{2}\Big(\hat{\sigma}_j^x \hat{\sigma}_{j+1}^{y}
- \hat{\sigma}_j^y \hat{\sigma}_{j+1}^{x} \Big) 
+\Big(h_1(t)+(-1)^j h_2(t)\Big)\hat{\sigma}_j^z\Big].
\label{ham_spin}
\end{eqnarray}
\normalsize
Here, $\hat{\sigma}^\alpha, \alpha=x,y,z$ are Pauli matrices and 
the parameters $J$ and $D$ represent the strengths of the nearest neighbor exchange couplings and DM interaction respectively, while $\gamma$  $(\neq 0)$ is the $x-y$ anisotropy in the exchange interaction. Note that the external magnetic field has site-dependent strengths, $h_j=h_1+(-1)^j h_2$, with  $j$ being the site index. We assume periodic boundary condition (PBC), i.e.\ $\sigma_{N+1}=\sigma_1$. We abbreviate the quantum spin model represented  by the above Hamiltonian as the DATXY model.

It is noteworthy to mention that, in this section as well as in Secs. \ref{sec_phases} and \ref{sec:DMent}, we  consider time-independent magnetic fields, $h_1$ and $h_2$, to study the properties of the system in equilibrium. Later, in Sec. \ref{sec_dyn}, we will consider time-dependent case (see, Eq. \eqref{eq:quench}) to examine dynamical behaviors of the DATXY model.

The Hamiltonian in Eq. (\ref{ham_spin})
can be mapped onto a two-component Fermi gas of spinless fermions on an 1D  lattice consisting of two sublattices $a$ and $b$, via Jordan-Wigner transformation, as \cite{c_cyclic}
\begin{eqnarray}
\hat{H} &=& \frac{J}{2} \sum_{j=1}^{N/2} \Big[(1 + i d)(\hat{a}^{\dagger}_{2j-1} \hat{b}_{2j} + \hat{b}^{\dagger}_{2j} \hat{a}_{2j+1} ) \nonumber \\
&+& (1 - i d) (\hat{b}^{\dagger}_{2j} \hat{a}_{2j-1} + \hat{a}^{\dagger}_{2j+1} \hat{b}^{\dagger}_{2j}) \nonumber \\
&+& \gamma \big(\hat{a}^{\dagger}_{2j-1}\hat{b}^{\dagger}_{2j} + \hat{b}^{\dagger}_{2j} \hat{a}^{\dagger}_{2j+1}
+ \hat{b}_{2j} \hat{a}_{2j-1} + \hat{a}_{2j+1} \hat{b}^{\dagger}_{2j}\big) \nonumber \\
&+& 2(\lambda_1 + \lambda_2) \hat{b}^{\dagger}_{2j} \hat{b}_{2j} + 2(\lambda_1 - \lambda_2) \hat{a}^{\dagger}_{2j-1} \hat{a}_{2j-1} - \lambda_1 \Big], \nonumber \\
\label{ham_ferm}
\end{eqnarray} 
where we define the dimensionless parameters as $d = D /J$, $\lambda_1 = h_1 / J$, and $\lambda_2 = h_2 / J$. 
Note that the existence of the two types of magnetic field (uniform and alternating) in the original model 
leads to the two sublattices in the fermionic model,   thereby resulting in 
two types of fermionic operators, $a^{\dagger}$ and $b^{\dagger}$.
The Hamiltonian in Eq. (\ref{ham_ferm}) can  again be simplified as
 $\hat{H}=\sum_{p=1}^{N/4}\hat{H}_p$, with 
\begin{eqnarray}
 \hat{H}_p &=& J \Big[ (\cos{\phi_p} +d \sin{\phi_p})(\hat{a}_p^{\dagger}\hat{b}_p + \hat{b}_p^{\dagger}\hat{a}_p) \nonumber \\
 &+& (\cos{\phi_p} - d \sin{\phi_p})(\hat{a}_{-p}^{\dagger}\hat{b}_{-p} + \hat{b}_{-p}^{\dagger}\hat{a}_{-p}) \nonumber \\
&-& i\gamma\sin{\phi_p}\big(\hat{a}_p^{\dagger}\hat{b}_{-p}^{\dagger}+\hat{a}_{p}\hat{b}_{-p}-\hat{a}_{-p}^{\dagger}\hat{b}_p^{\dagger}-\hat{a}_{-p}\hat{b}_{p}\big)\nonumber\\
 &+&\lambda_+(\hat{b}_p^{\dagger}\hat{b}_p+\hat{b}_{-p}^{\dagger}\hat{b}_{-p})
 +\lambda_-(\hat{a}_p^{\dagger}\hat{a}_p+\hat{a}_{-p}^{\dagger}\hat{a}_{-p})- 2\lambda_1 \Big]\nonumber \\
 \label{hp}
\end{eqnarray}
via Fourier transformations, where $\phi_p=2\pi p/N$, $\lambda_\pm=\lambda_1 \pm \lambda_2$, and $a_p^\dagger$ ($b_p^\dagger$) is fermionic operator in the momentum space. 
Since $[\hat{H}_p,\hat{H}_{p^\prime}]=0$, the above Fourier transformation decomposes the space, upon which $\hat{H}$ acts on, into non-interacting subspaces, each  having a dimension 16. Therefore, the spectrum of the Hamiltonian $\hat{H}$ can be obtained by performing diagonalization 
of $\hat{H}_p$, acting on the $p^{th}$ subspace. In Appendix \ref{ap:diagonalization}, we prescribe the method to diagonalize  $\hat{H}_p$ in details.

Furthermore, we can get two-site nearest-neighbor local density matrix of the canonical equilibrium state, corresponding to an ``even-odd" pair of spins,  as 
\begin{eqnarray}
\hat{\rho}_{eo} &=& \frac{1}{4} \Big[ \mathbb{I}_e \otimes \mathbb{I}_o + m_e^z \hat{\sigma}^z_e \otimes \mathbb{I}_o + m^z_o \mathbb{I}_e \otimes \hat{\sigma}_o^z \nonumber \\
&+& \sum_{\alpha = x,y,z} C^{\alpha \alpha} \hat{\sigma}_e^{\alpha} \otimes \hat{\sigma}_o^{\alpha} \nonumber \\
&+& C^{xy} \hat{\sigma}_e^{x} \otimes \hat{\sigma}_o^{y} + C^{yx} \hat{\sigma}_e^{y} \otimes \hat{\sigma}_o^{x} \Big],
\label{rho1}
\end{eqnarray}
using the same procedure mentioned in \cite{atxy_pra}. Here, the suffix $e$ ($o$) corresponds to the \textit{even} (\textit{odd}) lattice sites, $m^z_{e(o)} = \mbox{Tr}[\hat{\sigma}^z_{e(o)} \hat{\rho}_{eo}]$ is the magnetization in the $z$ direction, while $C^{\alpha \mu} = \mbox{Tr}[\hat{\sigma}_e^{\alpha} \otimes \hat{\sigma}_o^{\mu} \hat{\rho}_{eo}]$ are the two-site spin correlators. By using translational invariance, the correlator and magnetization operators  are defined respectively  as 
\begin{eqnarray}
\hat{C}^{\alpha\mu} &=& \frac{2}{N}\sum_{j = 1}^{N/2}\hat{\sigma}_{2j}^{\alpha}\hat{\sigma}_{2j+1}^{\mu}, \nonumber \\  
\hat{m}^{z}_o &=& \frac{2}{N}\sum_{j = 1}^{N/2}\hat{\sigma}_{2j - 1}^{z}, \nonumber \\
\hat{m}^{z}_e &=& \frac{2}{N}\sum_{j = 1}^{N/2}\hat{\sigma}_{2j}^{z},
\label{eq:corr_op}
\end{eqnarray}
where $\alpha, \mu = x,y$. Just like the Hamiltonian, successive applications of Jordan-Wigner and Fourier transformation block-diagonalize the above operators into different momentum subspaces as $\hat{C}^{\alpha\mu} = \frac{2}{N}\sum_{p = 1}^{N/4} \hat{C}^{\alpha\mu}_p$ and $\hat{m}^z_{e(o)} = \frac{2}{N}\sum_{p = 1}^{N/4} \hat{m}^z_{e(o),p}$.
We get the values of correlators as $C^{\alpha\mu} = \frac{2}{N}\sum_{p=1}^{N/4} C^{\alpha\mu}_p$, where
\small
\begin{eqnarray}
C^{\alpha\mu}_p = \frac{1}{Z_p}\text{Tr}[\hat{C}^{\alpha\mu}_p \exp(-\beta \hat{H}_p)], \ Z_p=\text{Tr}[\exp(-\beta\hat{H}_p)].
\label{eq:alt_cc}
\end{eqnarray}\normalsize
In the above equation, \(\exp(-\beta \hat{H}_p)\) is the canonical equilibrium (unnormalized) state with $\beta = 1/ {k_B T}$, $T$ and $k_B$ being the  temperature of the system and the Boltzmann constant respectively, and \(Z_p\) is the corresponding partition function. The procedure is similar for magnetizations. The matrix forms of $\hat{C}^{\alpha\mu}_p$ and $\hat{m}^z_{e(o),p}$ are given in Appendix \ref{ap:corr_fun}. Note that the $zz$-correlator, $C^{zz}$, can be obtained
using Wick's theorem as
\begin{eqnarray}
C^{zz} = m^z_e m^z_o - C^{xx} C^{yy} + C^{xy}C^{yx}.
\end{eqnarray}
To study the properties of the zero-temperature state of \(\hat{H}\), we simply choose $\beta\rightarrow\infty$. 

To obtain analytical expressions of the correlators, the magnetization and other physical quantities which are functions of them, we have to diagonalize matrices of dimension $16$, for which closed analytical forms are hard to obtain  in terms of all the parameters involved in the system. Hence,  identifications of different phases in this system requires numerical diagonalization of $16\times16$ matrix of $\hat{H}_p$.
The situation becomes much simpler if we turn off the alternating part of the field, i.e., $h_2=0$. In that case, $\hat{H}_p$ is a matrix of dimension $4$ which in turn can be further reduced to three sub-blocks of dimensions 2, 1, and 1, leading to compact analytical forms of magnetization and two-site spin correlators,  discussed in the next subsection.

\subsection{Uniform field case}
\label{uniformfield}

We now consider the UXY model in presence of the DM interaction (DUXY), i.e., $h_2 = 0$ in Eq. (\ref{ham_spin}). 
Instead of considering two different sublattices,  we can map the Hamiltonian
into a single component Fermi gas by the Jordon-Wigner transformation involving only one type of fermionic operator \cite{lsm,bm1,bm2}, 
$\hat{c}$, as   \cite{c_cyclic}
\begin{eqnarray}
\hat{H} &=& \frac{J}{2}\sum_{j = 1}^N \Big[ (1 +i d)\hat{c}_j^\dagger \hat{c}_{j+1}  + (1  -i d) \hat{c}_{j+1}^\dagger \hat{c}_j
\nonumber \\ 
 &+& \gamma  (\hat{c}_j^\dagger \hat{c}_{j+1}^\dagger + \hat{c}_{j+1} \hat{c}_{j})
+   \lambda_1\big( 2 \hat{c}_j^\dagger  \hat{c}_j  - 1\big)\Big].
\label{H_fermi_uniform}
\end{eqnarray}
Similar to the previous scenario,
the  Fourier transformation enables us to write
 $\hat{H} = \sum_{p=1}^{N/2} \hat{H}_p$, where the matrix form of $\hat{H}_p$ in the basis $\lbrace|0\rangle, \hat{c}_p^\dagger \hat{c}_{-p}^\dagger |0\rangle, \hat{c}_p^\dagger  |0\rangle, \hat{c}_{-p}^\dagger |0\rangle\ \rbrace$, is given by
\begin{equation}
\resizebox{.95\hsize}{!}{$\hat{H}_p = J \begin{bmatrix}
    -\lambda_1 & i\gamma \sin\phi_p & 0 & 0  \\
    - i\gamma \sin\phi_p & \lambda_1 + 2\cos\phi_p & 0 & 0 \\
    0 & 0 & \cos\phi_p + d\sin\phi_p & 0 \\
    0 & 0 & 0 & \cos\phi_p - d\sin\phi_p
  \end{bmatrix},$}
\label{eq:hp_matrix_uni}
\end{equation} \normalsize
with $\phi_p = 2\pi  p /N$. Note that due to the DM interaction, the matrix form of $\hat{H}_p$  changes only in the smaller sub-blocks. 
We can also compute the reduced two-site nearest-neighbor  density matrix between $n^{th}$ and $(n+1)^{th}$ lattice sites of the canonical equilibrium state as
\begin{eqnarray}
\hat{\rho}_{n,n+1} &=& \frac{1}{4} \Big[ \mathbb{I}_n \otimes \mathbb{I}_{n+1} + m^z \big(\hat{\sigma}^z_n \otimes \mathbb{I}_{n+1} + \mathbb{I}_n \otimes \hat{\sigma}_{n+1}^z \big) \nonumber \\
&+& \sum_{\alpha = x,y,z} C^{\alpha \alpha} \hat{\sigma}_{n}^{\alpha} \otimes \hat{\sigma}_{n+1}^{\alpha} \nonumber \\
&+& C^{xy} \hat{\sigma}_n^{x} \otimes \hat{\sigma}_{n+1}^{y} + C^{yx} \hat{\sigma}_n^{y} \otimes \hat{\sigma}_{n+1}^{x} \Big].
\label{rho2}
\end{eqnarray}
where $C^{\alpha\mu}$ and $m^z$ can be defined in a similar fashion as in Eqs. \eqref{eq:corr_op} and \eqref{eq:alt_cc} (see Appendix \ref{ap:corr_uni} for details).
In the thermodynamic limit ($N \rightarrow \infty$), the correlators and magnetizations of the zero-temperature state, i.e., with $\beta\rightarrow\infty$, are given in Table \ref{t11} (the same for thermal equilibrium state are given in Appendix \ref{ap:corr_uni}). 
Note that similar calculations for the DUXY model have been carried out in \cite{dm5}, where analytic forms of different structure factors of the model are derived, and their behaviors are explored.

\begin{table*}
\begin{tabular}{ |c|c| } 
 \hline
 Classical correlators & Analytical expressions   \\
  and magnetization & \\ 
 \hline
 & \\
  &  $ \frac{1}{\pi} \int_{0}^{\pi} d\phi_p \frac{1}{\Lambda_p} \left( - \gamma  \sin^2\phi_p + (\Lambda_p -  \cos\phi_p - \lambda_1)\cos\phi_p \right)$,  for $d < \gamma$ \\ 
 $C^{xx}$ & \\
  &   $ \frac{1}{\pi}[\int_{0}^{\phi_1} + \int_{\phi_2}^{\pi}] d\phi_p \frac{1}{\Lambda_p} \left( - \gamma  \sin^2\phi_p + (\Lambda_p -  \cos\phi_p - \lambda_1)\cos\phi_p \right) + \frac{1}{\pi}(\sin\phi_2 - \sin\phi_1)$, for $d>\gamma$  \\ 
  & \\
 \hline
 &\\
 &  $ \frac{1}{\pi} \int_{0}^{\pi} d\phi_p \frac{1}{\Lambda_p} \left(  \gamma  \sin^2\phi_p + (\Lambda_p -  \cos\phi_p - \lambda_1)\cos\phi_p \right)$, for $d < \gamma$ \\ 
 $C^{yy}$ & \\
  &   $ \frac{1}{\pi}[\int_{0}^{\phi_1} + \int_{\phi_2}^{\pi}] d\phi_p \frac{1}{\Lambda_p} \left(  \gamma  \sin^2\phi_p + (\Lambda_p - \cos\phi_p - \lambda_1)\cos\phi_p \right) + \frac{1}{\pi}(\sin\phi_2 - \sin\phi_1)$, for $d>\gamma$  \\ 
  & \\
  \hline
  & \\
  & $0$, for $d < \gamma$ \\
  $C^{xy}$ & \\
  & $\frac{1}{\pi}(\cos\phi_2 - \cos\phi_1)$, for $d>\gamma$ \\
  & \\ 
  \hline
  & \\
  & $0$, for $d < \gamma$ \\
  $C^{yx}$ & \\
  & $\frac{1}{\pi}(\cos\phi_1 - \cos\phi_2)$, for $d>\gamma$ \\
  & \\
  \hline
  & \\
  & $- \frac{1}{\pi}\int_{0}^\pi d\phi_p \frac{1}{\Lambda_p}(\lambda_1 + \cos\phi_p)$, for $d < \gamma$  \\
  $m^z$ & \\
  & $-\frac{1}{\pi}[\int_{0}^{\phi_1} + \int_{\phi_2}^\pi] d\phi_p \frac{1}{\Lambda_p}(\lambda_1 + \cos\phi_p)$, for $d>\gamma$ \\
  &\\
  \hline
  \end{tabular}
  \caption{Analytical expressions of classical correlators and magnetization of the zero-temperature state for the UXY model with DM interaction (DUXY). The expression of $\Lambda_p$ is given by $\sqrt{(\cos\phi_p + \lambda_1)^2 + \gamma^2\sin^2\phi_p}$,
while the expressions of $\phi_1$ and $\phi_2$ are given  \cite{phi_def}.  
   The expressions in the case of $d>\gamma$ are only true for real solutions of $(\phi_1,\phi_2)$. Otherwise, even in the case of $d>\gamma$, the $d < \gamma$ solution holds. Note that for $d = \gamma$, both the cases yield same expressions. The correlators and the magnetization of the thermal state are given in Appendix \ref{ap:corr_uni}.
   }
  \label{t11}
\end{table*}

The Hamiltonian, $\hat{H}_p$, given in Eq. (\ref{eq:hp_matrix_uni}), can be written as $\hat{H}_p = \hat{C}_p^{\dagger} \tilde{{H}}_p \hat{C}_p$, with $\hat{C}_p$ is the column vector, $(\hat{c}_p, \hat{c}_{-p}^{\dagger})$ and 
\begin{equation}
\resizebox{.9\hsize}{!}{$\tilde{{H}}_p = J
\begin{bmatrix}
\cos\phi_p + d \sin\phi_p + \lambda_1 & -i \gamma \sin\phi_p \\
i \gamma \sin\phi_p & - \cos\phi_p + d \sin\phi_p - \lambda_1
\end{bmatrix},$}
\end{equation}
where $\phi_p \in [0, \pi]$.
Now, the above matrix, ${\tilde{H}}_p$, has two eigenvalues, $J \big( d \sin\phi_p \pm \Lambda_p \ \big)$, with $\Lambda_p = \sqrt{(\cos\phi_p + \lambda_1)^2 + \gamma^2\sin^2\phi_p}$,
which give us the single-particle excitation spectrum of the model as
\begin{eqnarray}
\omega_{\phi_p} = J \big( d \sin\phi_p + \Lambda_p \ \big),
\label{eq:spectrum}
\end{eqnarray}
for $\phi_p \in [-\pi, \pi]$. For $0 \leq d < \gamma$, $\omega_{\phi_p}$ is always positive for any values of system parameters, and thus the ground state (or the zero-temperature state) of the model, in this scenario, is basically the vacuum of corresponding Bogoliubov operators. For the DUXY model, the Bogoliubov transformation does not depend on the DM interaction strength \cite{dm5}, and, as a consequence, the Bogoliubov vacuum remains independent of the value of $d$. In this scenario, the ground state energy also remains independent of $d$, as it comes from the upper $2 \times 2$ block of the Hamiltonian given in Eq. \eqref{eq:hp_matrix_uni}.
Therefore, we find that all the two-site correlators and the magnetization of the zero-temperature state of the model given in Table \ref{t11}, do not depend on the value of $d$ for $d < \gamma$.
However, for $d > \gamma$ and $\lambda_1^2 < 1+d^2 - \gamma^2$, $\omega_{\phi_p}$ becomes negative in the range $-\phi_2 < \phi_p < -\phi_1$ ($\phi_1$ and $\phi_2$ are mentioned in Table \ref{t11}), and the ground state is no longer the Bogoliubov vacuum, as in this case 
the modes in between $-\phi_2$ and $-\phi_1$ have to be filled to construct the ground state, which now depends on the value of $d$.
 
These results allow us to arrive at the following theorem.

\noindent 
\emph{\bf Theorem 1.} {\it For weak DM interaction strength~($0\le d < \gamma$), the zero-temperature state of the DUXY model is insensitive towards the DM interaction.} 

\begin{figure}
\includegraphics[width=0.7\linewidth]{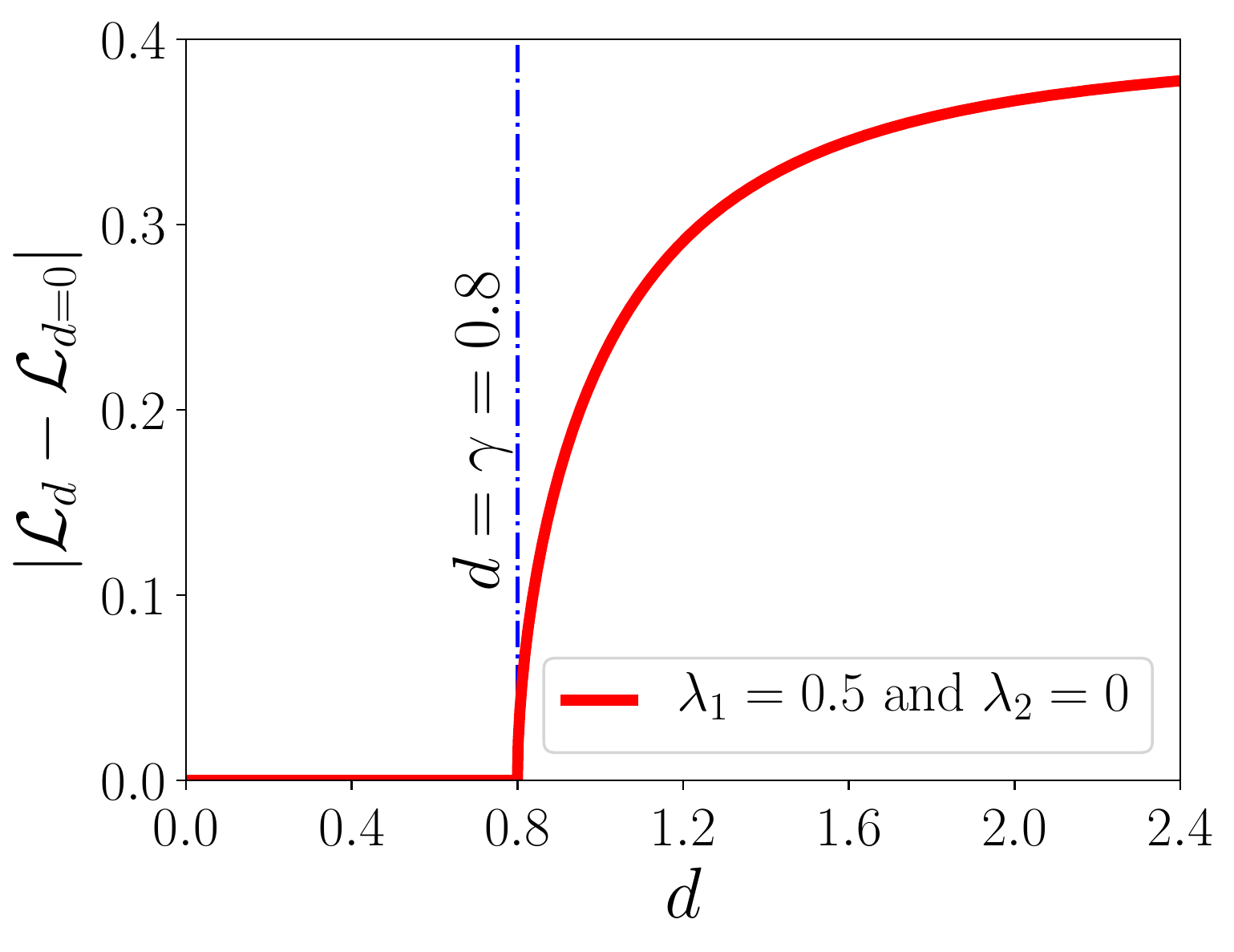}
\caption{(Color online.)
The absolute difference of nearest-neighbor entanglements, $|\mathcal{L}_{d} - \mathcal{L}_{d=0}|$, of the zero-temperature state of the DUXY model as a function of $d$.
Plot shows the insensitivity of the two-site entanglement of the zero-temperature state with the introduction of DM interaction, when $d < \gamma$.  We have shown that such behavior can be seen for all other physical quantities of this model. 
Here,  the zero-temperature state is computed using 
calculation explained in Sec. \ref{sec_model}.
Both the axes are dimensionless.
We set \(\gamma =0.8\). Unless otherwise stated,  we choose $\gamma = 0.8$ for depictions throughout this paper. Note that
the results reported here are independent of the values of \(\gamma\). 
}
\label{fig:insen}
\end{figure}

We highlight this insensitivity by plotting 
the absolute difference of nearest-neighbor entanglements, quantified by logarithmic negativity (LN) \cite{neg1, neg2, neg3, neg4, neg5} (see Sec. \ref{sec:DMent} for definition), for $d = 0$ and $d > 0$ as a function of $d$ in Fig. \ref{fig:insen}. It is also interesting to note that the correlators and the magnetization are insensitive even for $d > \gamma$, when $\phi_1$ and $\phi_2$ \cite{phi_def}, mentioned in Table \ref{t11}, do not have real solutions, i.e., for $\lambda_1^2 > 1+d^2 - \gamma^2$.
 However, the difference emerges for a canonical equilibrium state with finite temperature as shown in the succeeding sections. This is so because only the zero-temperature state of the DUXY model does not depend on the DM interaction strength for $d<\gamma$, but the excited states do, and therefore thermal excitations in the finite temperature scenario incorporate the effects of the DM term. We observe these characteristics from our analytical analysis of the DUXY model.  However, the exact physical reason behind these features is still elusive to us, and is yet to be explored.


With this formalism in hand, we are now ready to investigate the phase boundaries of the quantum DATXY chain. 

\section{Phase boundaries at zero temperature}
\label{sec_phases}

In this section, we find out and detect  different phases of the zero-temperature state of this model by using suitable order parameters, and identify the corresponding critical lines, along which quantum phase transitions occur, by investigating the signatures of energy gap closing. For investigation, we divide the parameter range into two sub-categories,  motivated by Table \ref{t11} -- {(1)} $0 \leq d < \gamma$ and {(2)} $d > \gamma$.

\subsection{Phase boundaries for weak DM interactions: $0 \leq d < \gamma$}

To discuss  the phases of $\hat{H}$ with $0 \leq d < \gamma$, we first notice that 
the Hamiltonian, $\hat{H}_p$, given in Eq. (\ref{hp}) can be written as $\hat{H}_p = \hat{A}_p^{\dagger}  \tilde{H}_p \hat{A}_p$, where $\hat{A}_p$ is the column vector,  $(\hat{a}_p, \hat{b}_p, \hat{a}_{-p}^{\dagger}, \hat{b}_{-p}^{\dagger})$, and the $4 \times 4$ matrix, $\tilde{H}_p$, is given as
\begin{widetext}
\small
\begin{equation}
\tilde{H}_p = J
\begin{bmatrix}
(\lambda_1 - \lambda_2) &  (\cos\phi_p + d \sin\phi_p) & 0 & -i \gamma \sin\phi_p \\
(\cos\phi_p + d \sin\phi_p)  & (\lambda_1 + \lambda_2) & -i \gamma \sin\phi_p &0 \\
0 & i\gamma\sin\phi_p & -(\lambda_1 - \lambda_2) &  -(\cos\phi_p - d \sin\phi_p) \\
i \gamma \sin\phi_p & 0 & -(\cos\phi_p - d \sin\phi_p) & -(\lambda_1 + \lambda_2)
\end{bmatrix},
\label{eq:hp_matrix_alt}
\end{equation}
\normalsize
\end{widetext}
with $\phi_p \in [-\pi/2,\pi/2]$.
The above matrix has four eigenvalues, $\omega_{\phi_p}^k$ with $k = 1,2, 3, 4$, that satisfy $\omega_{\phi_p}^{1(2)} = - \omega_{-\phi_p}^{3(4)}$ for $\phi_p \in [0, \pi/2]$ \cite{ph_sym}. Therefore, we get the two bands of the
single-particle excitation spectrum as $\{\omega_{\phi_p}^1, \omega_{\phi_p}^2\}$ for 
$\phi_p \in [-\pi/2, \pi/2]$.
We detect the quantum critical lines by tracking  the gap closing  in the $(\lambda_1,\lambda_2)$-plane in the thermodynamic limit  as identified by the vanishing excitation energies ($\mbox{min}_{p, k=1,2} \{|\omega^k_{\phi_p}|\} \rightarrow 0$).
For $d < \gamma$, substantial numerical search reveals that $\mbox{min}_{p, k=1,2} \{|\omega^k_{\phi_p}|\}$ may possess
vanishingly small value only when $\phi_p = 0$ or $\pm \pi/2$.
For $\phi_p = 0$, we obtain
\begin{eqnarray}
\omega^1_0 &=& J \big(\lambda_1 + \sqrt{1 + \lambda_2^2}\big),  \nonumber \\
\omega^2_0 &=& J \big(\lambda_1 - \sqrt{1 + \lambda_2^2}\big).
\end{eqnarray}
Clearly, we get $\omega^2_0 = 0$ when $\lambda_1 = \sqrt{1 + \lambda^2_2}$, while $\omega^1_0 = 0$ for $\lambda_1 = -\sqrt{1 + \lambda^2_2}$, 
thereby 
implying a gapless line, 
\begin{eqnarray}
\lambda_1^2 = 1 + \lambda^2_2,
\label{eq:critical_line_1}
\end{eqnarray} 
which indicates a quantum phase transition. Notice that the above critical line does not depend on the value of $d$. Next, for $\phi_p = \pm \pi /2$, we get $\{\omega_{\pm \pi/2}^k\}$ as 
\small \begin{eqnarray}
\omega_{\pm \pi/2}^1 &=& J\big(\sqrt{\lambda_1^2 + \gamma^2} + \sqrt{\lambda_2^2 + d^2}\big), 
\nonumber \\
\omega_{\pm \pi/2}^2 &=& J\big(\sqrt{\lambda_1^2 + \gamma^2} - \sqrt{\lambda_2^2 + d^2}\big).
\end{eqnarray}
\normalsize
It is easy to check that $\omega_{\pm \pi/2}^2 = 0$ for $\lambda^2_2 = \lambda_1^2 + \gamma^2 - d^2$, giving another phase boundary as
\begin{eqnarray}
\lambda^2_2 = \lambda_1^2 + \gamma^2 - d^2, 
\label{eq:critical_line_2}
\end{eqnarray}
which depends both on anisotropy and DM interaction parameters. Eqs. (\ref{eq:critical_line_1}) and (\ref{eq:critical_line_2}) indicate that the XY model in presence of  DM interaction along with uniform and alternating transverse magnetic fields posses rich phase diagram and hence
it will be interesting to  characterize  the quantum phases present in this model, using appropriate order parameters which we will do in next subsections.

\begin{figure}
\includegraphics[width=0.9\linewidth]{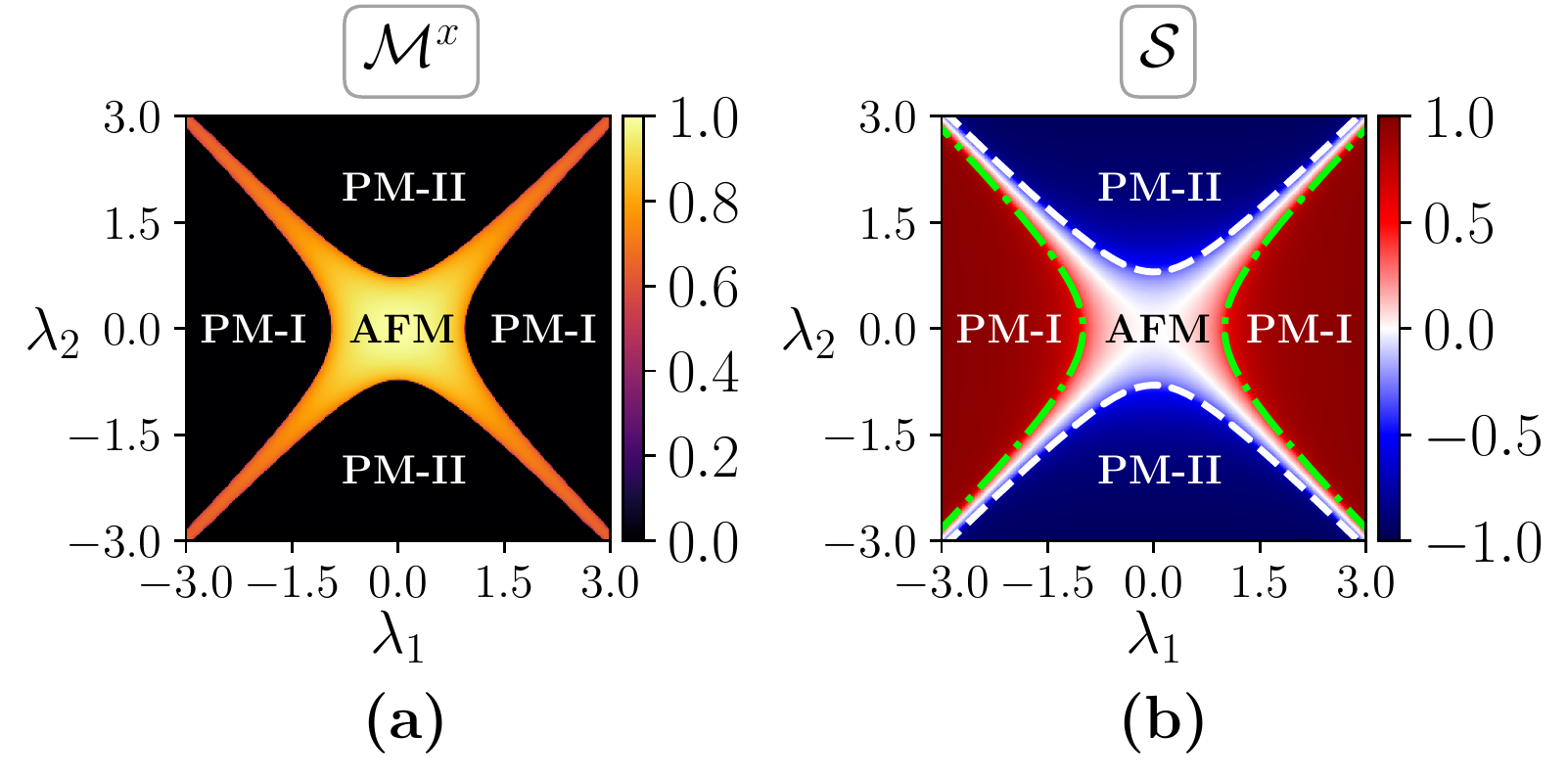}
\caption{(Color online.) Characterization of phases  of the ATXY model ($d = 0$) in the thermodynamic limit. (a) We plot the antiferromagnetic order parameter, $\mathcal{M}^x$, in the $(\lambda_1,\lambda_2)$-plane, which is non-zero in the AFM phase and zero in other two phases.
(b) The magnetization in the $z$-direction has staggered order in the PM-II phase, whereas in PM-I phase, it is ordered. Therefore, $\mathcal{S} \equiv m^z_e m^z_o$ has high negative value in PM-II phase, while it possesses high positive value in PM-I phase.
 To calculate $\mathcal{M}^x$ of the zero temperature state, we use DMRG with $N=100$, while to obtain $\mathcal{S}$, we use analytical methods explained in Secs. \ref{sec_model} and \ref{sec_phases}.
Both the axes are dimensionless.
}
\label{fig:d_0}
\end{figure}

\begin{figure*}
\includegraphics[width=0.85\linewidth]{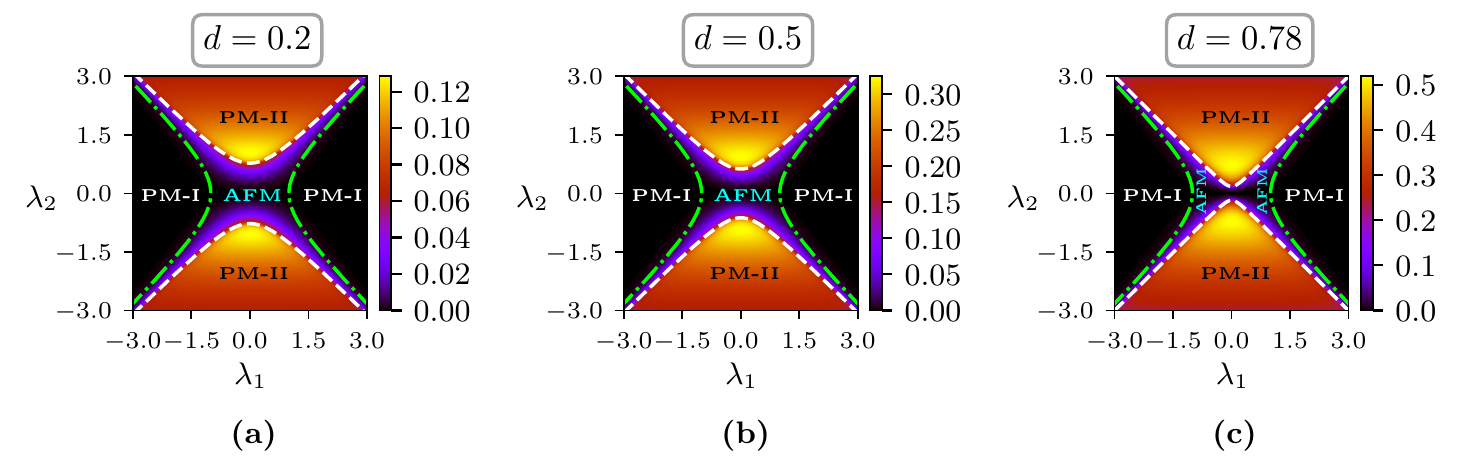}
\caption{(Color online.) The chiral order parameter, $\mathcal{C}$, of the zero-temperature state (obtained in Secs. \ref{sec_model} and \ref{sec_phases}) for different values of $d < \gamma = 0.8$
 in the $(\lambda_1, \lambda_2)$-plane. For $d > 0$, chiral order is developed in the PM-II phase and in some regions of AFM phase, while in PM-I phase, the order parameter remains zero. 
Both the axes are dimensionless and all the other system parameters are same as in Fig. \ref{fig:d_0}.}
\label{fig:chiral_dlg}
\end{figure*}

\begin{figure*}
\includegraphics[width=0.85\linewidth]{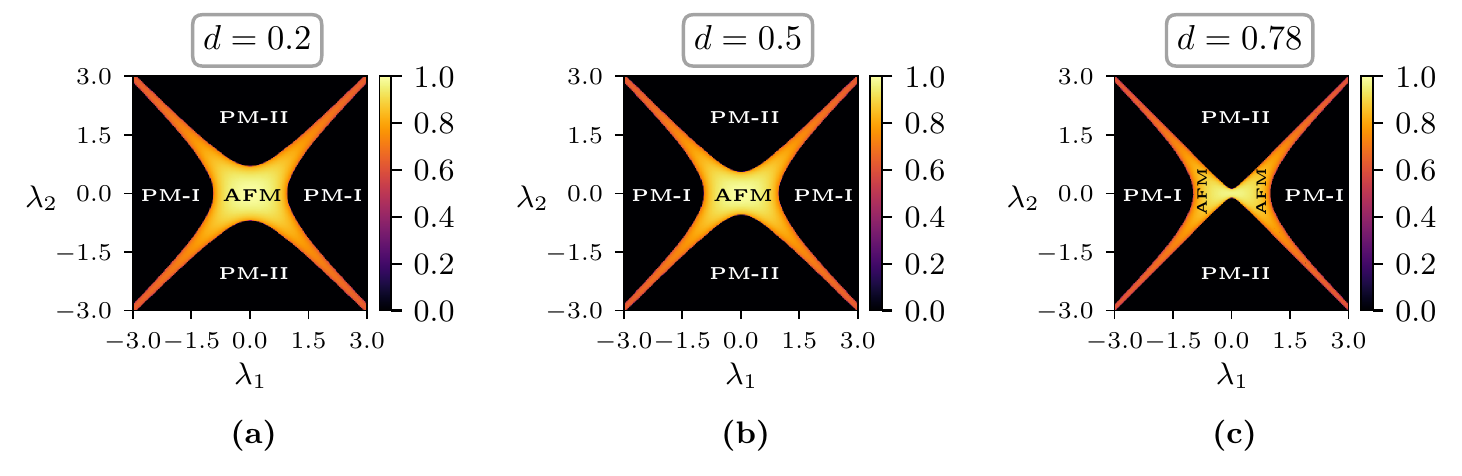}
\caption{(Color online.) The antiferromagnetic order parameter, $\mathcal{M}^x$, of the zero-temperature state of the 
DATXY model for different values of $d < \gamma = 0.8$
 in the $(\lambda_1, \lambda_2)$-plane. The zero-temperature state is calculated using DMRG with $N=100$.
Both the axes are dimensionless.}
\label{fig:mag_dlg}
\end{figure*}

\begin{figure}
\includegraphics[width=0.85\linewidth]{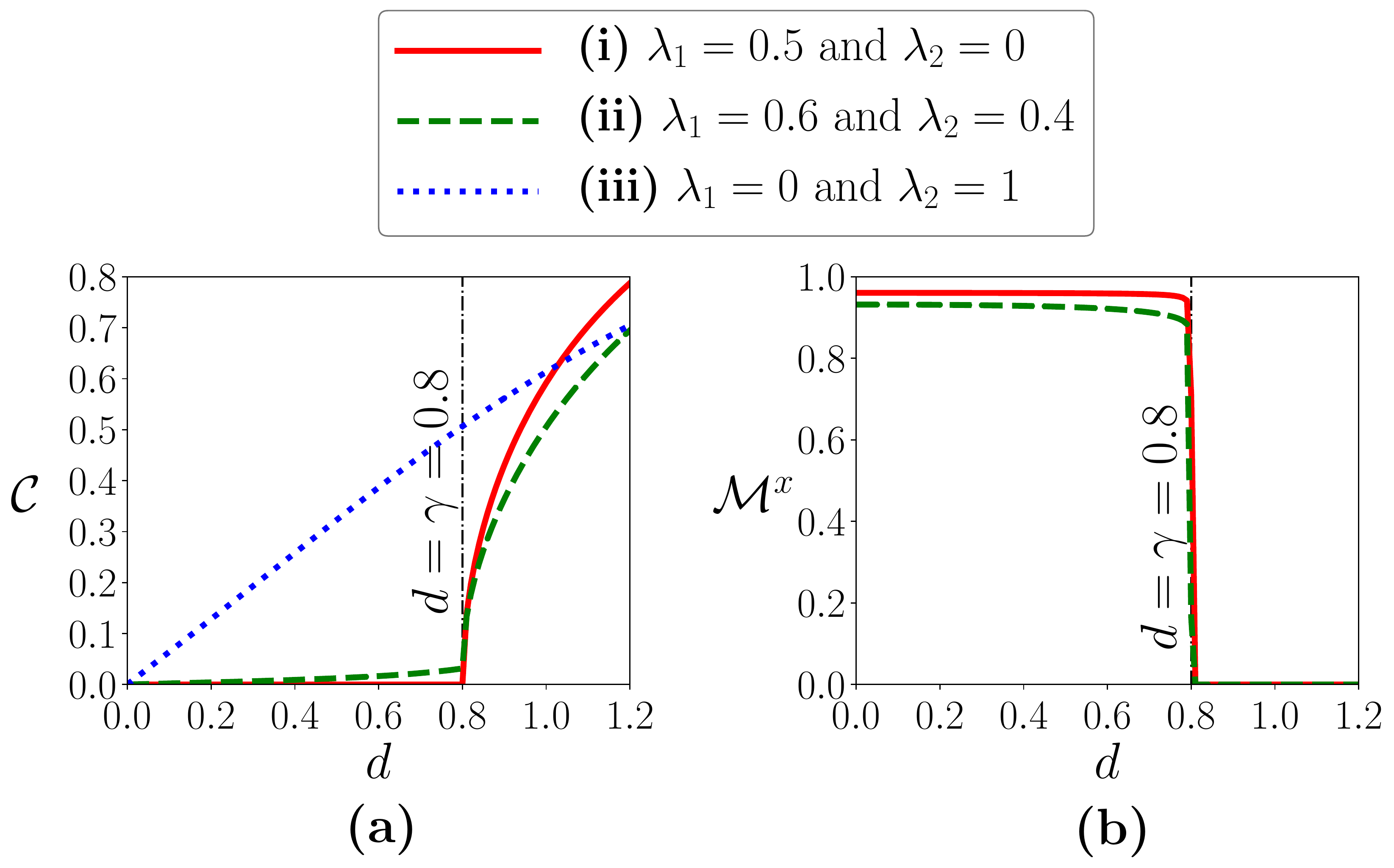}
\caption{(Color online.) Quantum phase transition at $d = \gamma$, as detected by the order parameters $\mathcal{C}$ and $\mathcal{M}^x$. Both the axes in (a) and (b) are dimensionless.
(a) Development of chiral order with the DM interaction strength, $d$. We plot the chiral order parameter, $\mathcal{C}$, as a function of $d$ for three different values of $(\lambda_1, \lambda_2)$-pair. The AFM phase transforms into the CH phase for $d > \gamma$, which is indicated by the sharp changes in $\mathcal{C}$ at $d = \gamma$ (red solid and green dashed lines).  
PM-II phase develops chiral order for $d > 0$ (blue dotted line), which is just an artifact of non-zero $d$.
(b) The antiferromagnetic order parameter, $\mathcal{M}^x$, against the strength of DM interaction, $d$, 
for two different values of $(\lambda_1, \lambda_2)$-pair. $\mathcal{M}^x$ vanishes for $d > \gamma$, as the AFM phase transforms into the CH phase. All the other considerations are same as in Fig. \ref{fig:d_0}.
All the axes are dimensionless.
}
\label{fig:chiral_2d}
\end{figure}

\begin{figure*}
\includegraphics[width=0.75\linewidth]{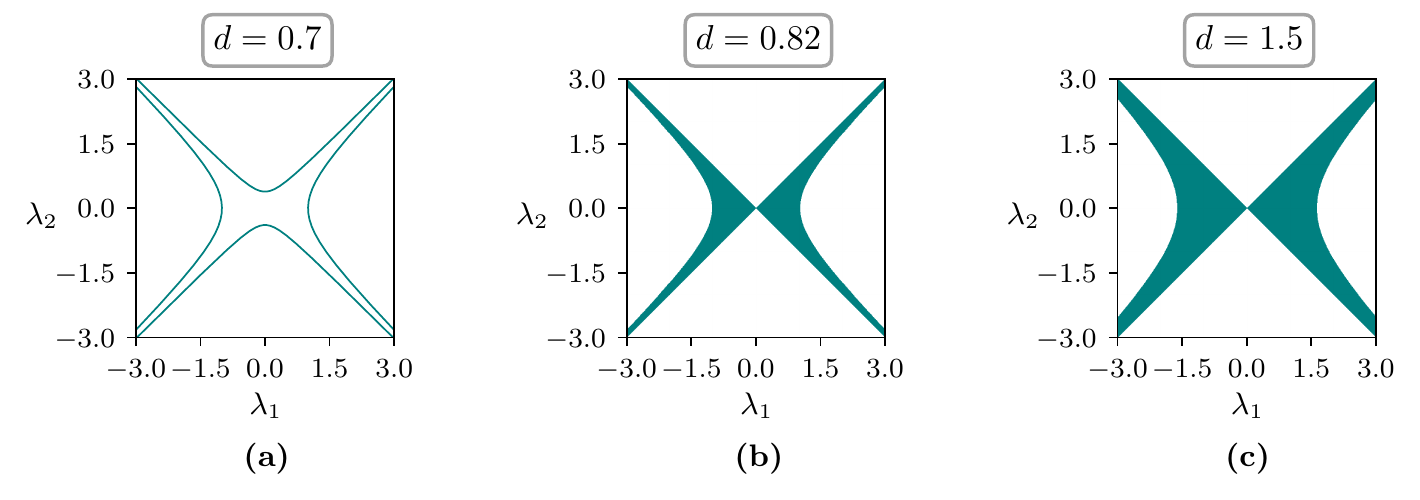}
\caption{(Color online.) Regions in the $(\lambda_1, \lambda_2)$-plane where 
minimum excitation energy, $\mbox{min}_{p, k=1,2}  \{|\omega^k_{\phi_p}|\}$, that is required to drive the ground state to the first excited state, is vanishingly small. Clearly, for $d < \gamma = 0.8$, we obtain only quantum critical lines (AFM $\leftrightarrow$ PM-I and AFM $\leftrightarrow$ PM-II) as  gapless, while for $d > \gamma$, the entire CH phase is gapless as explained in Secs. \ref{sec_model} and \ref{sec_phases}. 
Both the axes are dimensionless.}
\label{fig:gapless}
\end{figure*}

\begin{figure*}
\includegraphics[width=0.85\linewidth]{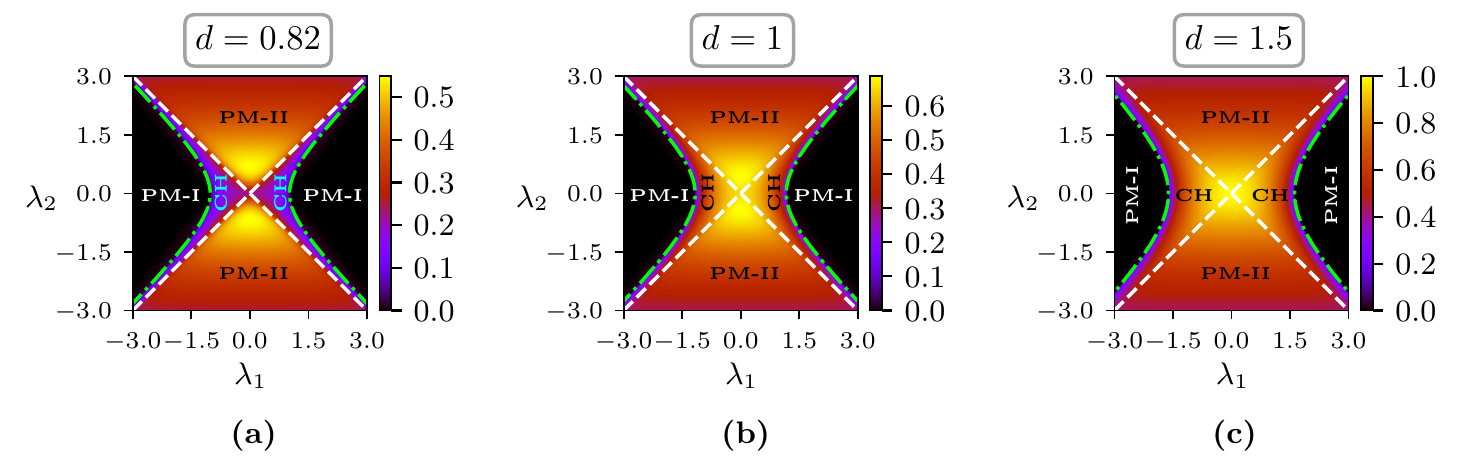}
\caption{(Color online.) The chiral order parameter,  $\mathcal{C}$, for different values of $d > \gamma = 0.8$ in the $(\lambda_1, \lambda_2)$-plane. Clearly, for $d > \gamma$, chiral order is developed in the former AFM phase. 
All the other considerations are same as in Fig. \ref{fig:d_0}.
Both the axes are dimensionless.}
\label{fig:chiral_dgg}
\end{figure*}

\subsubsection{Characterization of phases for $d = 0$}

Before discussing the scenario with non-zero $d$, let us discuss the model with $d=0$, which corresponds to the ATXY model \cite{atxy_pra},
having three different quantum phases, namely, (1) anti-ferromagnetic (AFM), (2) paramagnetic I (PM-I), and (3) paramagnetic II (PM-II) phases \cite{atxy_pra}. These three phases correspond to three distinct types of orders
-- (i) AFM phase has staggered magnetic order in the $(x,y)$-plane; (ii)  in PM-I phase, $\langle\sigma^z_j\rangle = m^z_j$ is uniformly ordered in the $z$-direction, and (iii) $m_j^z$ has a staggered order in PM-II phase \cite{dmtopmii}.

To distinguish AFM  from the PM phases, we add a small alternating field, $h_x$, in the $x$ direction of magnitude $10^{-8}$, in units of $J$, to the original Hamiltonian,  so that the new Hamiltonian reads as
\begin{eqnarray}
\hat{H}' = \hat{H} + h_x \sum_{j = 1}^N (-1)^j \hat{\sigma}^x_j.
\label{eq:dmrg_ham}
\end{eqnarray}
Note that the above Hamiltonian can not be diagonalized analytically, and hence 
we use density-matrix renormalization group (DMRG) technique \cite{dmrg1, dmrg2, dmrg3, dmrg_ex1, dmrg_ex2, dmrg_ex3} to obtain the zero-temperature state for $N=100$ with open boundary condition. To identify the antiferromagnetic order, we examine the order parameter, staggered magnetization in the $x$ direction, $\mathcal{M}^x$, defined as 
\begin{eqnarray}
\mathcal{M}^x = \Big|\frac{1}{N} \sum_{j = 1}^N (-1)^j \langle \hat{\sigma}^x_j \rangle \Big| = \Big|\frac{1}{N} \sum_{j = 1}^N (-1)^j m^x_j \Big|.
\end{eqnarray}
Fig. \ref{fig:d_0} (a) depicts the value of $\mathcal{M}^x$ in the $(\lambda_1, \lambda_2)$-plane. Clearly, for $\lambda_1^2 < 1 + \lambda^2_2$ and $\lambda_2^2 < \lambda_1^2 + \gamma^2$, the order parameter $\mathcal{M}^x$ is non-vanishing,  indicating the AFM phase.

As mentioned earlier,
in PM-I phase, 
the quantity $\mathcal{S} \equiv m^z_j m^z_{j+1} \equiv m^z_e m^z_o$  possess high positive value, since $m^z_j$'s are uniformly ordered in the $z$-direction, while  $\mathcal{S}$ have high negative value in PM-II, as shown in 
 Fig. \ref{fig:d_0} (b). 
 We observe that for $\lambda_1^2 > 1 + \lambda_2^2$, $\mathcal{S}$ has high positive value,  faithfully characterizing the  PM-I phase, while for $\lambda_2^2 > \lambda_1^2 + \gamma^2$, it has high negative value, thereby signaling the PM-II phase.

\subsubsection{Characterization of phases for $0 < d < \gamma$}

Let us now move to non-zero values of $d < \gamma$. As mentioned earlier,  the AFM $\leftrightarrow$ PM-I transition line, given in Eq. (\ref{eq:critical_line_1}), remains same, while AFM $\leftrightarrow$ PM-II critical line (Eq. (\ref{eq:critical_line_2})) get modified with the presence of $d$. 
Specifically,  a new type of order, the chiral order, gets developed in some regions  of the $(\lambda_1, \lambda_2)$-plane which can be demonstrated by considering the physical quantity,  known as chiral order parameter 
\cite{PRB38}, given by
\begin{eqnarray}
\mathcal{C} = \Big|\frac{1}{N}\sum_{i=1}^{N}\langle \sigma^x_j\sigma^y_{j+1} - \sigma^y_j\sigma^x_{j+1} \rangle \Big| = |C^{xy} - C^{yx}|.
\label{chiral_order}
\end{eqnarray}
As depicted in Fig. \ref{fig:chiral_dlg}, we find that  in the zero-temperature state, $\mathcal{C} $ possess non-vanishing values 
in the PM-II phase and in some regions of the AFM phase, close to the PM-II phase while  it vanishes in PM-I.
It is important to stress here  that due to the DM interaction,  chiral order is created, leading to  $\mathcal{C} \neq 0$, although  no non-analyticity or discontinuity found around $d = 0$, thereby signaling  the absence of quantum phase transition with $d \rightarrow 0$. 

We observe that this case also the antiferromagnetic order parameter, $\mathcal{M}^x$, rightfully characterizes the AFM phase in the presence of non-zero DM interaction (see Fig. \ref{fig:mag_dlg}).  
We will show in the following subsection that situations will change as soon as $d > \gamma$, and 
such changes will lead to a quantum phase transition at $d = \gamma$.

\textbf{Note:} As mentioned earlier, the chiral order parameter, $\mathcal{C}$, is also non-vanishing in the regions of the AFM phase, which are close to the PM-II phase (see Fig. \ref{fig:chiral_dlg}). In these regions, where both $\mathcal{M}^x$ and $\mathcal{C}$ possess finite values, the antiferromagnetic state has an incommensurate order, as opposed to the regions with vanishing $\mathcal{C}$, where we have commensurate antiferromagnet. However, there is no quantum criticality between between antiferromagnetic regions with non-zero $\mathcal{C}$ and vanishing $\mathcal{C}$. This makes it quite difficult to differentiate between these two regions in $(\lambda_1, \lambda_2, \gamma)$-space analytically. However, it is clearly understandable from Fig. \ref{fig:chiral_dlg} (and from the expressions 
given in Table \ref{t11})
 that the alternating magnetic field, $h_2$ (or $\lambda_2= h_2/J$), must be non-zero for stabilizing 
incommensurate antiferromagnetic order.

Summarizing, in the case of  \( 0 < d < \gamma \), we show that   there exists two critical lines, namely
 \begin{eqnarray}
\lambda_1^2 &=& 1 + \lambda_2^2 \ \ \ \ \ \ \ \ \ \  \textrm{(AFM} \leftrightarrow \textrm{PM-I)}, \nonumber \\
\lambda_2^2 &=& \lambda_1^2 + \gamma^2 -d^2 \ \ \ \textrm{(AFM} \leftrightarrow \textrm{PM-II)}.
\label{eq:critical_line_3}
\end{eqnarray}
On top of that, a new chiral order emerges which will be  prominent when DM interaction dominates over the anisotropy parameter, as we will discuss in the next subsection.

\subsection{Phase boundaries for strong DM interactions: $d > \gamma$}
\label{subsec_DMorder}

Let us now study  quantum phases of the zero-temperature state when $d > \gamma$, i.e., when the second term in Eq. (\ref{ham_spin}) dominates over the first one. We will show that certain phase emerges in this situation due to trade-off between  $d$ and $ \gamma$ which has already been seen in $d < \gamma\). 
For a fixed values of \(\lambda_1\) and \(\lambda_2\), we demonstrate a phase transition at $d = \gamma$ in Fig. \ref{fig:chiral_2d}, using the order parameters $\mathcal{C}$ and $\mathcal{M}^x$. For demonstration,  we choose three kinds of parameter values -- Case-(i)  the DUXY model in the  AFM phase; Case-(ii) the AFM phase of the DATXY model; Case-(iii) the PM-II phase of the DATXY model.  
Let us discuss the observations  
from Fig. \ref{fig:chiral_2d} (a).
\begin{enumerate}
\item In the Case-(i), chiral order parameter  remains zero for non-zero $d < \gamma$ and
  sharply becomes non-zero at \(d= \gamma\). 
  \item In Case-(ii),  \(\mathcal{C}\)  becomes non-vanishing  immediately after  the introduction of $d$  although there is a sharp increase of \(\mathcal{C}\) at \( d =\gamma\).  
    \item In contrast, $\mathcal{C}$  is  a smooth increasing function in the entire range of $d$ in the PM-II phase (Case-(iii)).  
\end{enumerate}   
The first two observations suggest that there is a  quantum phase transition  on the onset of \(d=\gamma\) from AFM to a chiral ordered (CH) phase. Existence of this quantum criticality was also hinted earlier in Fig. \ref{fig:insen}, where bipartite nearest-neighbor entanglement of the zero-temperature state shows a sharp change at $d = \gamma$.
We will later show analytically that there indeed exists a critical point at  \(d=\gamma\).  

 Such indication of change of phase can also be made if one studies  the behavior of $\mathcal{M}^x$.
In Fig. \ref{fig:chiral_2d} (b), the variation of the antiferromagnetic order parameter, $\mathcal{M}^x$, with $d$ is depicted,  when the systems are in the AFM phase  (Case-(i) and -(ii)). In both the scenarios, we find that  $\mathcal{M}^x$ posses  positive high values for $d < \gamma$, while it vanishes for $d > \gamma$. It clearly indicates that at the cost of destruction of the AFM order, a new phase, the CH phase, appears after \(d = \gamma\), which will be shown below by the following theorem.

\noindent\textbf{Theorem 2.}
\emph{A new gapless CH phase emerges in place of AFM phase for $d > \gamma$, in the DATXY model.}

\noindent\textbf{Proof.} 
%
Let us we first concentrate on the  DUXY model  (i.e., $\lambda_2 = 0$) for $d > \gamma$.
For $d > \gamma$ and $\lambda_1 \leq 1 + d^2 - \gamma^2$, 
the single-particle excitation spectrum, $\omega_{\phi_p}$, given in Eq. \eqref{eq:spectrum}, becomes zero at $\phi_p = -\phi_{1}, -\phi_{2}$, where $\phi_{1}$ and $\phi_2$ are mentioned in Table \ref{t11}, 
so that we have quasi-particle excitations with infinitesimal energy at $\phi_p =-\phi_1,  -\phi_2$, which renders the spectrum gapless.

If we now investigate the chiral order parameter, $\mathcal{C}$, we get from Table \ref{t11} that it is identically zero everywhere for $d < \gamma$, whereas
for $d > \gamma$,
we find that
\begin{eqnarray}
\mathcal{C}(\lambda_1) &=& \frac{2}{\pi}|\cos\phi_1 - \cos\phi_2|, \text{ for } \lambda_1^2 \leq 1+d^2-\gamma^2 \nonumber \\
&=& 0, \text{ for } \lambda_1^2> 1 + d^2-\gamma^2.
\label{dm_order}
\end{eqnarray}
This clearly shows that the chiral order gets developed in the new gapless phase, i.e., in the CH phase, while 
for $\lambda_1^2 \leq 1+d^2-\gamma^2$, the chiral order gets completely destroyed, and we are only left with a paramagnetic (PM-I) state.
Therefore, for $d > \gamma$, we obtain a quantum critical line as
\begin{eqnarray}
\lambda_1^2 = 1+d^2-\gamma^2  \ \ \ \ \textrm{(CH} \leftrightarrow \textrm{PM-I)},
\end{eqnarray}
for the DUXY model. 

Let us now focus on the DATXY model.
In presence of non-zero alternating field, i.e.,  $\lambda_2 \neq 0$,
following Eq. (\ref{eq:hp_matrix_alt}), we compute the minimum energy (i.e., $\mbox{min}_{p, k=1,2}  \{|\omega^k_{\phi_p}|\}$) required to excite the ground state to the first excited state for different values of $d$ in the thermodynamic limit to find out the gapless regions. Fig. \ref{fig:gapless} points out the regions in the $(\lambda_1, \lambda_2)$-plane where this minimum excitation energy is vanishingly small.
It is clear from the figure that, when $d < \gamma$, only quantum critical lines (i.e., Eq. (\ref{eq:critical_line_3})) can have gapless excitations, whereas for $d > \gamma$,  there exists indeed a \emph{gapless} phase in the $(\lambda_1,\lambda_2)$-plane, whose phase boundaries are found to be as follows:
\begin{eqnarray}
\lambda_1^2 &=& 1+ \lambda_2^2 + d^2-\gamma^2  \ \ \ \ \textrm{(CH} \leftrightarrow \textrm{PM-I)}, \nonumber \\
\lambda_1 &=& \pm \lambda_2\,\, \ \ \ \ \ \ \ \ \ \ \ \ \ \ \ \ \ \ \ \ \ \ \ \ \textrm{(CH} \leftrightarrow \textrm{PM-II)}.
\label{eq:critical_line_4}
\end{eqnarray}
This gapless  phase is found to be chiral in the DATXY model  by studying the chiral order parameter $\mathcal{C}$ (see Fig. \ref{fig:chiral_dgg}).
\hfill $\blacksquare$

\begin{figure}
\includegraphics[width=0.85\linewidth]{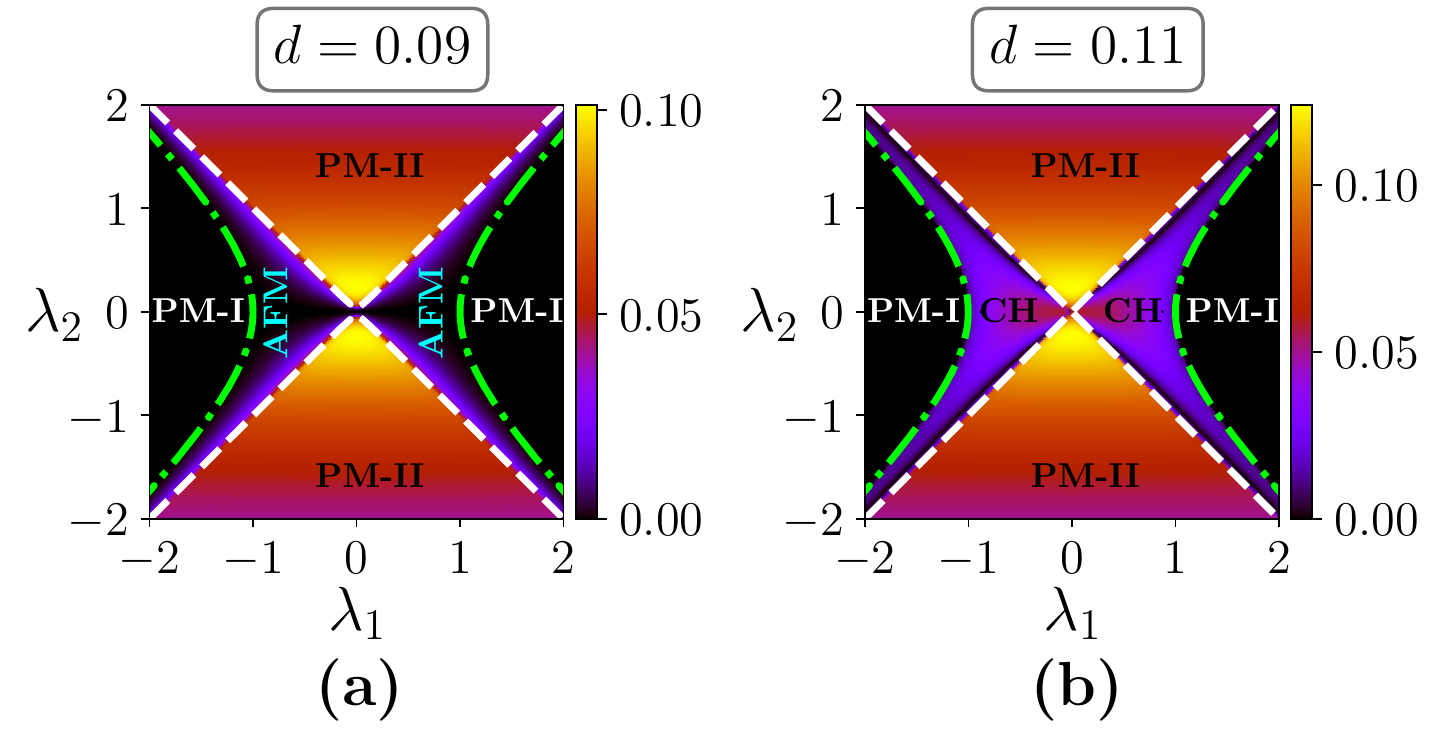}
\caption{(Color online.) The chiral order parameter,  $\mathcal{C}$, for different values of $d$ in the $(\lambda_1, \lambda_2)$-plane for $\gamma=0.1$. Clearly, for lower values of the anisotropy parameter, the onset of CH phase occurs for lower values of the DM interaction strength, $d$. 
All the other considerations are same as in Fig. \ref{fig:d_0}.
The zero-temperature state is obtained as explained in Secs. \ref{sec_model} and \ref{sec_phases}.
Both the axes are dimensionless.
}
\label{fig:chiral_g0.1}
\end{figure}

\begin{figure*}
\includegraphics[width=0.85\linewidth]{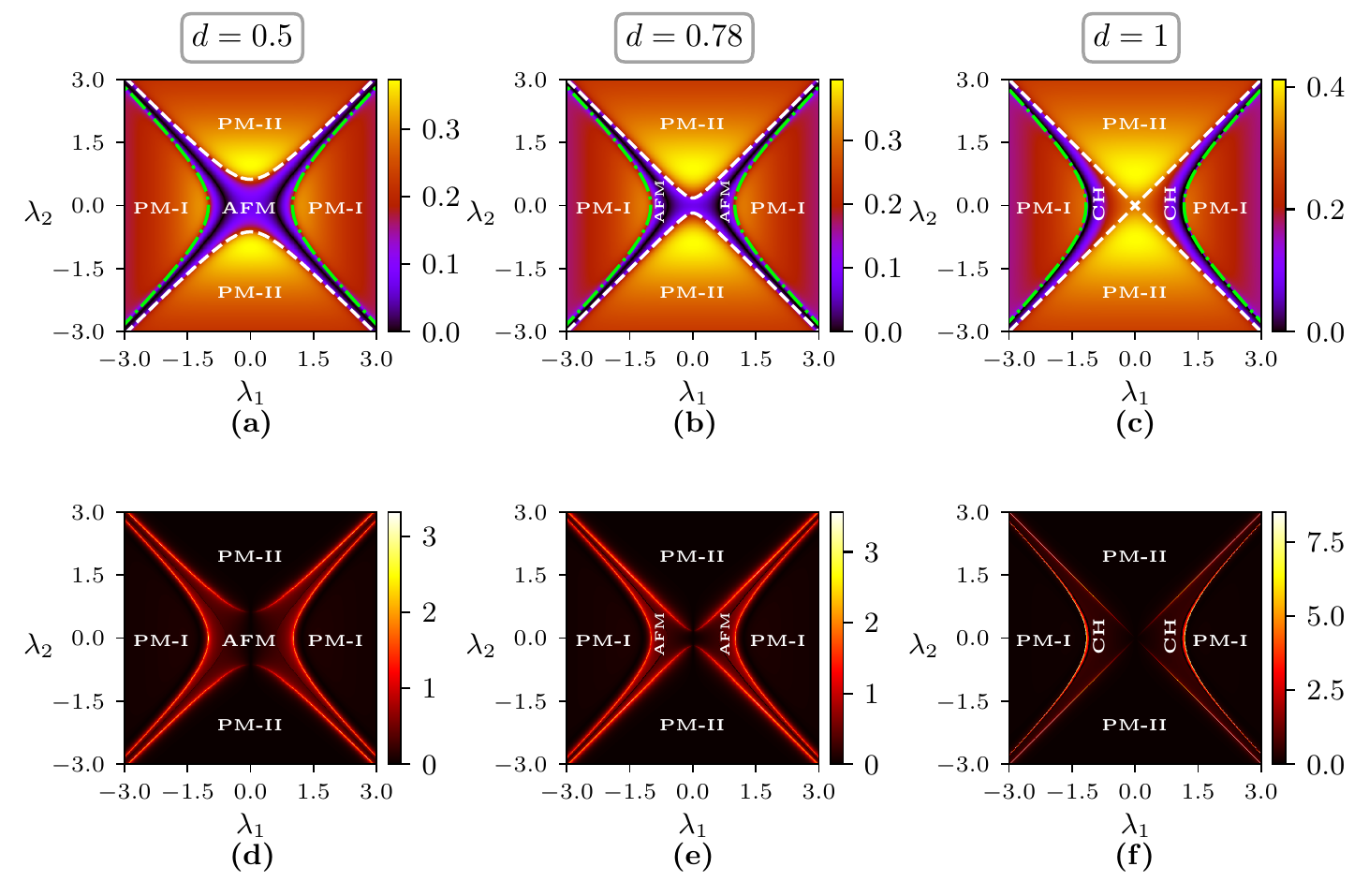}
\caption{(Color online.) Nearest-neighbor two-site entanglement (first row) and its first derivative with respect to $\lambda_1$ (second row) of the zero-temperature state in \((\lambda_1, \lambda_2)\)-plane by using the method obtained in Secs. \ref{sec_model} and \ref{sec_phases}. Each column corresponds to different choices of $d$. Values of $d$ for the left two columns are chosen in a way that it is  strictly less than \(\gamma\), while the right one corresponds to a value of $d$,  strictly greater than \(\gamma\). Both the axes are dimensionless. 
Black backgrounds in (d)-(f) represent the finite value in the derivative of entanglement with respect to $\lambda_1$ while the red lines depict the divergence of $|\frac{\partial  \mathcal{L}}{\partial \lambda_1}|$ through the lines.
Similar features can also be observed when derivatives are taken with respect to $\lambda_2$.
All the axes are dimensionless.}
\label{fig:ent}
\end{figure*}

It is interesting to note that the critical line, given in Eq. (\ref{eq:critical_line_2}), becomes $\lambda_1 = \pm \lambda_2$ for  $d = \gamma$ and remains same for higher values of $d$,
 which can not be explained by using Eq. (\ref{eq:critical_line_2}). 
 
Note that here we have taken $\gamma =0.8$ for a demonstrative purpose and it turns out that in this case, some of the expression yields nice numerical values. 
However, our analysis is valid for all $\gamma \in (0,1]$ and $d  \geq 0$. Therefore, if the value of the anisotropy parameter is low, say 0.1, we can get CH phase even for a very low strength of the DM interaction i.e.,  $d>0.1$ (see Fig. \ref{fig:chiral_g0.1}), which may be much easier to achieve experimentally \cite{dmf1, dmf2, dmf3, dmf4}.


\begin{figure}
\includegraphics[width=\linewidth]{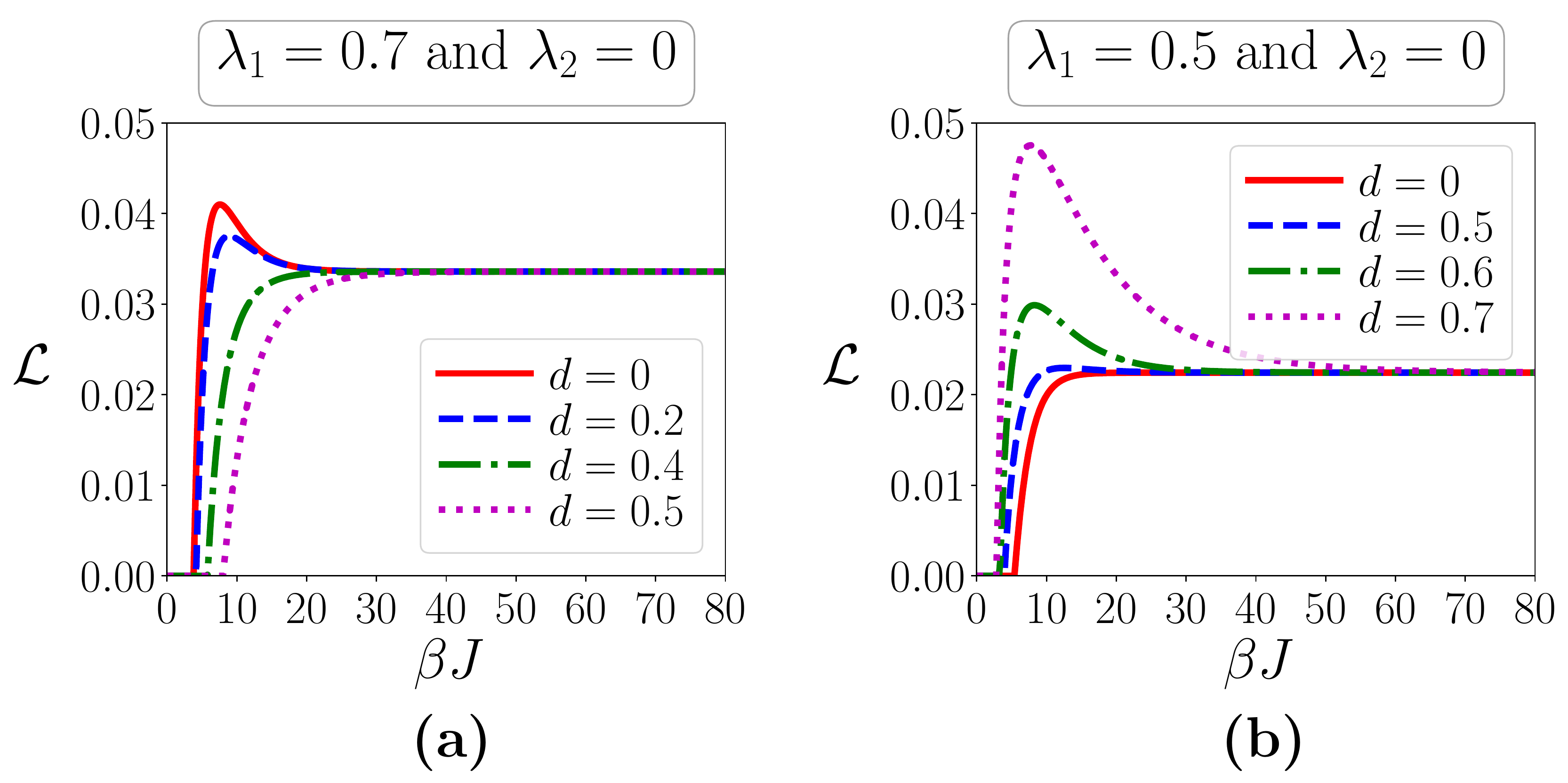}
\caption{(Color online.) Logarithmic Negativity vs. $\beta J$ for different values of $d$ of the canonical equilibrium state of the DUXY model (i.e., \(\lambda_2 =0\)) as in Secs. \ref{sec_model} and \ref{sec_phases}.  In (a), \(\lambda_1\) is chosen in a way that it is non-monotonic for $d=0$. With the increase of $d$, it slowly becomes monotonic with temperature. (b) shows the opposite feature.
Note that for high values of $\beta J$, all of them converge to a same value which confirms the analytical result that the zero-temperature state is insensitive to DM interaction, when $d < \gamma$. 
Both the axes are dimensionless.}
\label{fig:LN_beta11}
\end{figure}

\begin{figure*}
\includegraphics[width=0.9\linewidth]{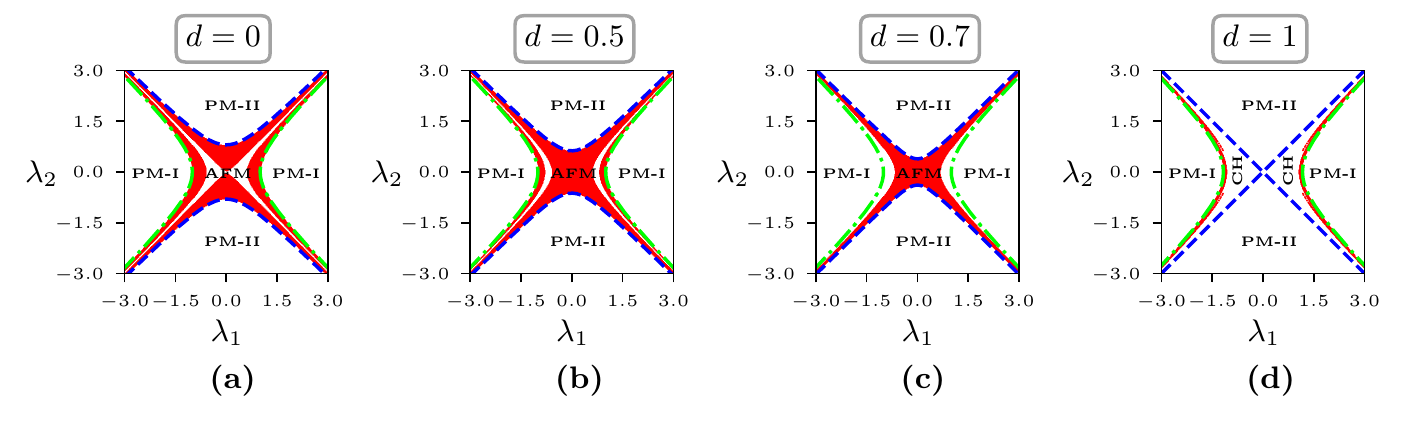}
\caption{(Color online.) Map of non-monotonic region  of LN against $\beta J$ for different values of $d$ in the \((\lambda_1, \lambda_2)\)-plane. (a), (b) and (c) represent $d$ which are strictly less than $\gamma$ while (d) is for $d> \gamma$. 
The canonical equilibrium state is computed using 
calculation explained in Sec. \ref{sec_model}.
Both the axes are dimensionless.}
\label{fig:LN_beta11_map}
\end{figure*}
 
 \section{Detection of phase boundaries by entanglement}
\label{sec:DMent}

In this section, we will demonstrate that the first derivatives of  quantum entanglement of nearest-neighbor two-site reduced density matrix of the zero-temperature state of the DATXY model  can detect all the critical lines discussed above.  We also explore the behavior of the nearest-neighbor entanglement, as measured by logarithmic negativity (LN) \cite{neg1, neg2, neg3, neg4, neg5},  in the thermodynamic limit.

Before that, let us give the definition of LN, which requires the concept of negativity \cite{neg1, neg2, neg3, neg4, neg5}, another measure of bipartite entanglement.
The negativity \cite{neg1, neg2, neg3, neg4, neg5}, \({\cal N}(\rho_{AB})\), for a bipartite state \(\rho_{AB}\), is  the absolute value of 
the sum of all the negative eigenvalues of  the partial transposed state, \(\rho_{AB}^{T_{A}}\) of $\rho_{AB}$, with  partial transposition  being taken with respect to the subsystem \(A\) \cite{ppt1, ppt2}. 
 Mathematically, it is defined as 
 \begin{equation}
  {\cal N}(\rho_{AB})=\frac{\|\rho_{AB}^{T_A}\|_1-1}{2},
  \label{eq:negativity}
 \end{equation}
where $\|\rho\|_1 \equiv \mbox{tr}\sqrt{\rho^\dag \rho}$ is the trace-norm of  matrix $\rho$.
Finally, LN, defined in terms of negativity, is given by 
\begin{equation}
\mathcal{L}(\rho_{AB}) = \log_2 [2 {\cal N}(\rho_{AB}) + 1].
\label{eq:LN}
\end{equation}
Its positive value ensures that the state has nonvanishing bipartite entanglement.

  The exact computation of LN can be performed using the form of the nearest-neighbor density matrix of the zero-temperature state,  given in Eqs. (\ref{rho1}) and (\ref{rho2}).  In Fig. \ref{fig:ent} (a)-(c), we map the value of LN as a function of $\lambda_1$ and $\lambda_2$ for three different values of the DM interaction strength, $d$, namely $d = 0.5$, $0.78$, and $1.0$. For depiction, we choose $d$ values in such a way that $d=0.5$ and $0.78$ is less than $\gamma$ and $d=1 > \gamma$. The observations from the investigations of entanglement is as follows:

\begin{enumerate}
\item When $d < \gamma$, entanglements in PM-I and PM-II regions are higher than that  in the AFM phase and the region near boundaries between PM-II and AFM posses a high amount of entanglement. Interestingly, the two site entanglement pattern itself can identify the transitions by strikingly changing its values. 

\item For $d > \gamma$, when the AFM phase transforms to the CH phase, the trends of entanglement  changes drastically 
in the neighborhood of \(\lambda_1=\lambda_2 =0\), which is in CH. In particular, the low entanglement regions shift towards the PM-I $\rightarrow$ CH critical lines, and quite high amount of entangled states are created near $(0,0)$ point. Moreover, we notice that the entanglement content in this neighborhood, belonging to PM-II, is much higher compared to the other regions in \((\lambda_1, \lambda_2)\)-plane  with $d<\gamma$.
\end{enumerate}  
 Note here that in the AFM phase, like the model without DM interaction, we find that there  exist surfaces having vanishing entanglement which can be called as ``factorization surfaces". The effects of DM interaction on these surfaces will be discussed in the succeeding subsection.

Let us now see whether bipartite entanglement  can accurately signal the critical lines found in the preceeding section  \cite{AmicoRMP, ADP_Aditi} (cf. \cite{Area}).  
For such identification, we perform first derivatives of entanglements of the zero-temperature state, for example with respect to \(\lambda_1\) (see  second row of Fig.\ \ref{fig:ent}), and  map $| \partial{\mathcal{L}}/\partial{\lambda_1}|$ in the $(\lambda_1, \lambda_2)$-plane for the same choices of $d$. From the figure, it is clear that the first derivative of LN diverge at AFM $\leftrightarrow$ PM-I and AFM $\leftrightarrow$ PM-II boundaries for $d < \gamma$.  Interestingly, in  case of $d > \gamma$, the derivative also diverge on the onset of CH $\leftrightarrow$ PM-I and CH $\leftrightarrow$ PM-II transitions. 
Note that this is one of the first demonstration where bipartite entanglement can successfully identify a gapped-to-gapless phase transition. The similar feature can also be seen by considering $|\partial\mathcal{L}/ \partial\lambda_2|$.
Note that we have already demonstrated in Fig. \ref{fig:insen} that entanglement can also detect AFM $\leftrightarrow$ CH transition.
The results clearly establish that entanglement can accurately detect different types of quantum phases, gapped as well as gapless (cf. \cite{cfails1, cfails2}) and corresponding quantum critical lines, thereby establishing itself as a universal detector for identifying  phase boundaries in the DATXY model.

\subsection{Non-monotonic-to-monotonic transition in entanglement with temperature}

The absolute zero-temperature is not easy to reach in experiments. And so it is interesting to investigate the patterns of entanglement of the canonical equilibrium state, \( \hat{\rho}^{eq} = \exp (- \beta \hat{H} )/Z\)  with varying temperature as well as with $d$. It is expected that entanglement, bipartite as well as multipartite,  goes to zero, when  \(\beta \rightarrow 0\) since the state becomes maximally mixed while   it saturates to entanglement of the zero-temperature state with high values of \(\beta\). Apart from these extreme cases,   it was shown that  \cite{atxy_pra, nonmono1, nonmono2, nonmono3, nonmono4, nonmono5, amader_facto} that entanglement shows a counter-intuitive behavior with respect to temperature --  it increases with the increase of temperature for specific choices of system parameters -- phenomena known as non-monotonicity of entanglement with temperature.  The question is whether such non-monotonic (monotonic) nature of entanglement  persists in the presence of DM interaction. From the continuity argument, we can infer that the behavior remains same for  small values of $d$ which is also depicted in Fig. \ref{fig:LN_beta11}. Interestingly, it modifies its behavior qualitatively with the increase in the strength of $d$.
%

With the substantially high values of $d < \gamma$ and suitable choices of \(\lambda_1\),  and \(\lambda_2\), we observe that entanglement becomes monotonic  from its non-monotonic nature with  $\beta$  and vice-versa with the variation of $d$ (see Fig. \ref{fig:LN_beta11}). For example, in the DUXY model, we find that  there is a  non-monotonic  to monotonic transition in entanglement of the thermal equilibrium state with the increase of  \(d\) as depicted  Fig. \ref{fig:LN_beta11} (a), while the opposite is seen in (b). 
Figure \ref{fig:LN_beta11} also shows  that   the  entanglement of the thermal state of the DUXY model is no longer insensitive towards  $d$ at reasonably high temperature, which is not the case for moderately low temperature as well as for zero-temperature, confirming the result obtained in Fig. \ref{fig:insen}.
Such transition is also noticed in the AFM phase of the DATXY model as depicted in Fig. \ref{fig:LN_beta11_map}. 
However, in case of PM-II phase of the DATXY model, such transitions are absent with \(\beta\) for any positive values of $d$  (see Fig. \ref{fig:LN_beta11_map}). Specifically, the monotonic or non-monotonic characteristics of entanglement with temperature does not change even in  presence of $d$ in the PM-II phase. Moreover, for lower values of the anisotropy parameter, $\gamma$, such transitions can be seen in the PM-I also.
Therefore, the transition observed here crucially depends on the phases on which the system belongs as well as on the value of the anisotropy parameter. It is also interesting to note that, with high enough $d > \gamma$, 
except very small regions in the CH, almost all the regions in $(\lambda_1, \lambda_2)$-plane shows monotonic entanglement variation with $\beta$ (Fig. \ref{fig:LN_beta11_map} (d)).

\subsection{Factorization volumes} 
\label{subsec_FL}

\begin{figure}
\includegraphics[width=0.9\linewidth]{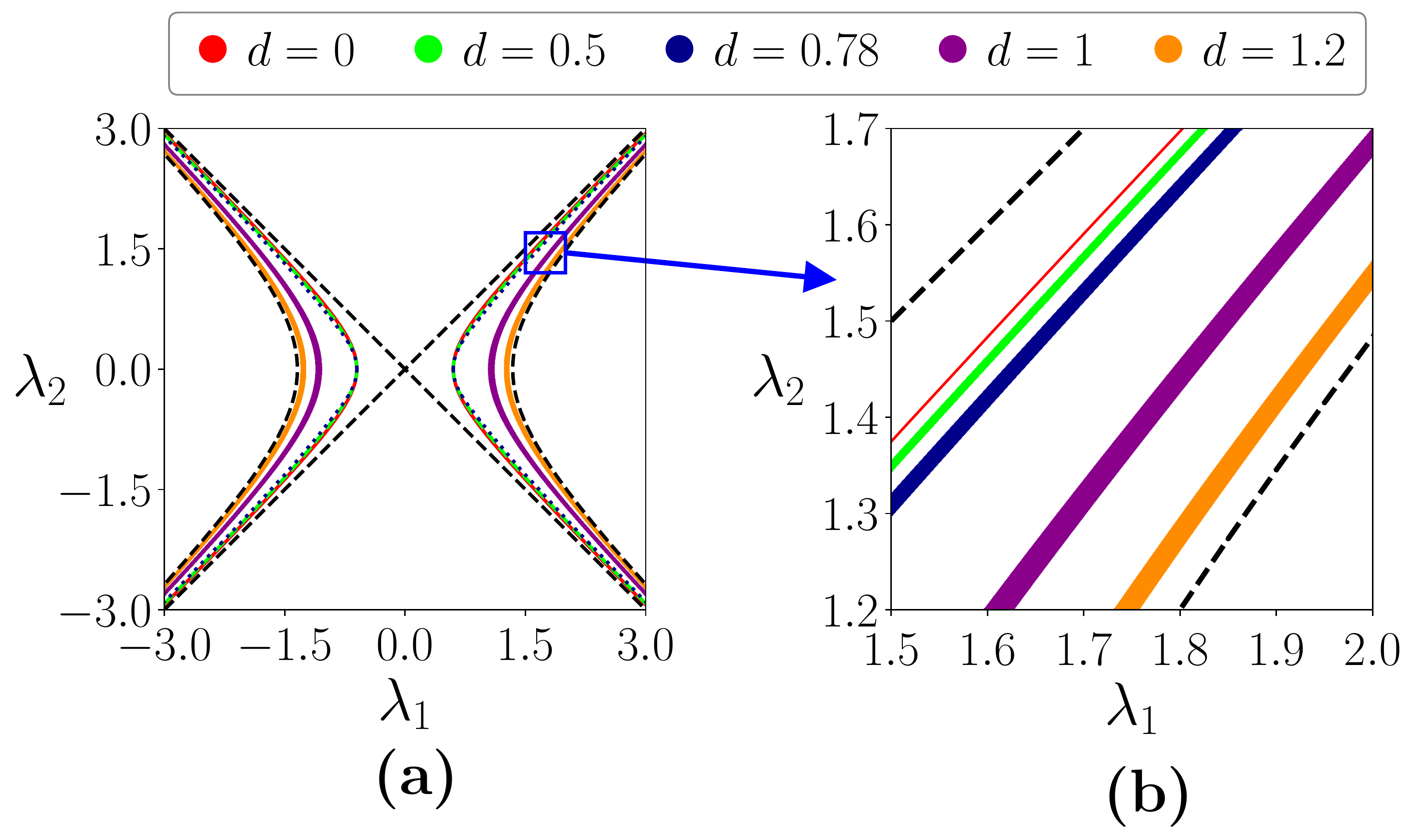}
\caption{(Color online.) The set of states having vanishing entanglement at zero-temperature in \((\lambda_1, \lambda_2)\)-plane, obtained in \ref{sec_model}, for different values of $d$ -- factorization volumes. The left figure shows that factorization volumes shift towards the PM-I $\leftrightarrow$ AFM/CH critical lines while the right one clearly shows that factorization surfaces become volume in presence of small amount of $d$ and the volume increases with increasing $d < \gamma$. The (black)  dashed  lines depict the phase boundaries for $d = 1.2$.
 Both the axes are dimensionless. }
\label{fig:LN_factor}
\end{figure}

In the AFM phase of the ATXY model,  the zero-temperature state contains  surfaces, given by 
\begin{equation}
\lambda_1^2 = \lambda_2^2 + 1 - \gamma^2,
\label{eq:factorize}
\end{equation}
which possess vanishing bipartite as well as multiprtite entanglement \cite{atxy_pra, amader_facto, facto1, facto2, facto3, facto4, facto5, facto6, facto7, facto8, facto9, facto10, facto11, facto12, facto13, facto14, facto15, facto16, facto17}. The state in this surface is doubly degenerate, and they are product across every bipartitions, thereby the name  ``factorization surfaces''. In the UXY model, i.e. when \(\lambda_2 =0\),  the surface reduces to a circle in \((\lambda_1-\gamma)\)-plane, which  represents a point  for fixed \(\gamma\)  with \(\lambda_1 = \pm \sqrt{1 - \gamma^2}\). 

At this point, it  is natural to ask whether there is any effects of DM interaction on these surfaces. 
We find here that in  presence of  DM interaction in the ATXY model,  factorization surfaces become  volumes at  zero-temperature except at the point corresponding to the DUXY model -- we call them factorization volumes. Specifically, we observe that till \(d < \gamma\), these surfaces shift slowly towards the boundary of PM-I and AFM  phases except in the DUXY model, and the volume, containing states with vanishing entanglment, increases with the strength of $d$. In contrast, when \(d > \gamma\),  the factorization volumes of the DUXY as well as the DATXY models deviate faster  towards PM-I $\leftrightarrow$ AFM critical lines in comparison with the case of \(d<\gamma\), as depicted in Figs. \ref{fig:LN_factor} and \ref{fig:ent}. Hence the results obtained here show that at zero-temperature, entanglement can be generated with the help of moderate DM interaction in the entire factorization volumes of the DATXY model at the cost of relocating the volumes towards the quantum critical lines (cf. \cite{amader_facto}).

\section{Sustainable Entanglement after switching on  DM interaction } 
\label{sec_dyn}

Upto now, we have studied the system either at zero or at finite temperatures. We now move to investigate the behavior of entanglement with time. When the evolution is governed by the time dependent version of the Hamiltonian, given by
\begin{eqnarray}
\hat{H}(t)=\frac{1}{2}\sum_{j = 1}^{N}\Big[J\Big\{\frac{1+\gamma}{2}\hat{\sigma}_j^x\hat{\sigma}_{j+1}^{x}+\frac{1-\gamma}{2}\hat{\sigma}_j^y\hat{\sigma}_{j+1}^{y}\Big\} \nonumber \\ 
+ \frac{D}{2}\big(\hat{\sigma}_j^x \hat{\sigma}_{j+1}^{y}
- \hat{\sigma}_j^y \hat{\sigma}_{j+1}^{x} \big) 
+\big(h_1(t)+(-1)^j h_2(t)\big)\hat{\sigma}_j^z\Big].\nonumber\\
\label{ham_spin_t}
\end{eqnarray}
Such study of dynamics of entanglement plays a crucial role in realization of quantum technology \cite{FP1, FP2, FP3}. 
Investigations are carried out by analyzing the response of the model with DM interaction to a sudden quench of both the uniform and the alternating parts of the magnetic field.
The quenching is performed as
\begin{eqnarray}
 h_1(t)= \left\{
 \begin{array}{cc}
 h_1, & t\leq 0  \\
 0, & t>0
\end{array}\right.,\;\;
 h_2(t)= \left\{
 \begin{array}{cc}
 h_2, & t\leq 0  \\
 0, & t>0
\end{array}\right..
\label{eq:quench}
\end{eqnarray}
Recently, similar sudden-quench of magnetic field has been considered in Ref\cite{pre_duxy}, where effects of DM interaction in the context of work distribution and the irreversible entropy production in the DUXY model have been studied. In this paper, we analyze the consequences of the presence of the DM interaction in the dynamics of entanglement in the DATXY model.

\begin{table*}
\begin{tabular}{ |c|c| } 
 \hline
 Classical correlators  & Analytical expressions   \\
 and magnetization & \\ 
 \hline
 & \\
  &  $  \int_{0}^{\pi} d\phi_p K_{-1}(\tilde{t},\phi_p) , \text{for } d < \gamma$ \\ 
 $C^{xx}(t)$ & \\
  &   $ [\int_{0}^{\phi_1} + \int_{\phi_2}^{\pi}] d\phi_p K_{-1}(\tilde{t},\phi_p)  , \text{for } d>\gamma$  \\ 
  & \\
 \hline
 &\\
 &  $ \int_{0}^{\pi} d\phi_p K_{1}(\tilde{t},\phi_p) , \text{for } d < \gamma$ \\ 
 $C^{yy}(t)$ & \\
  &   $ [\int_{0}^{\phi_1} + \int_{\phi_2}^{\pi}] d\phi_p K_{1}(\tilde{t},\phi_p)  , \text{for } d>\gamma$  \\ 
  & \\
  \hline
  &\\
  & $\int_{0}^{\pi} d\phi_p S(\tilde{t},\phi_p)$, for $d < \gamma$ \\
  $C^{xy}(t)$ & \\
  & $[\int_{0}^{\phi_1} + \int_{\phi_2}^\pi] d\phi_p S(\tilde{t},\phi_p) + \frac{1}{\pi}(\cos\phi_2 - \cos\phi_1)$, for $d>\gamma$ \\
&\\ 
  \hline
  &\\ 
  & $\int_{0}^{\pi} d\phi_p S(\tilde{t},\phi_p)$, for $d < \gamma$ \\
  $C^{yx}(t)$ & \\
  & $[\int_{0}^{\phi_1} + \int_{\phi_2}^\pi] d\phi_p S(\tilde{t},\phi_p) + \frac{1}{\pi}(\cos\phi_1 - \cos\phi_2)$, for $d>\gamma$ \\
&\\   
  \hline
  & \\
  & $\int_{0}^\pi d\phi_p M(\tilde{t},\phi_p)$, for $d < \gamma$  \\
  $m^z(t)$ & \\
  & $[\int_{0}^{\phi_1} + \int_{\phi_2}^\pi] d\phi_p M(\tilde{t},\phi_p)$, for $d>\gamma$ \\
  &\\
  \hline
  
  \end{tabular}
  \caption{Analytical expressions of time evolved classical correlators and magnetization for $t>0$ of the zero-temperature state after switching off the uniform field. The expressions in the case for $d>\gamma$ are only true for real solutions of $(\phi_1,\phi_2)$ \cite{phi_def}. Otherwise even in the case of $d>\gamma$, the $d<\gamma$ solution holds. Note that for $d = \gamma$, both the cases yield same expressions.}
  \label{t2}
\end{table*}

 \begin{figure}
\includegraphics[width=\linewidth]{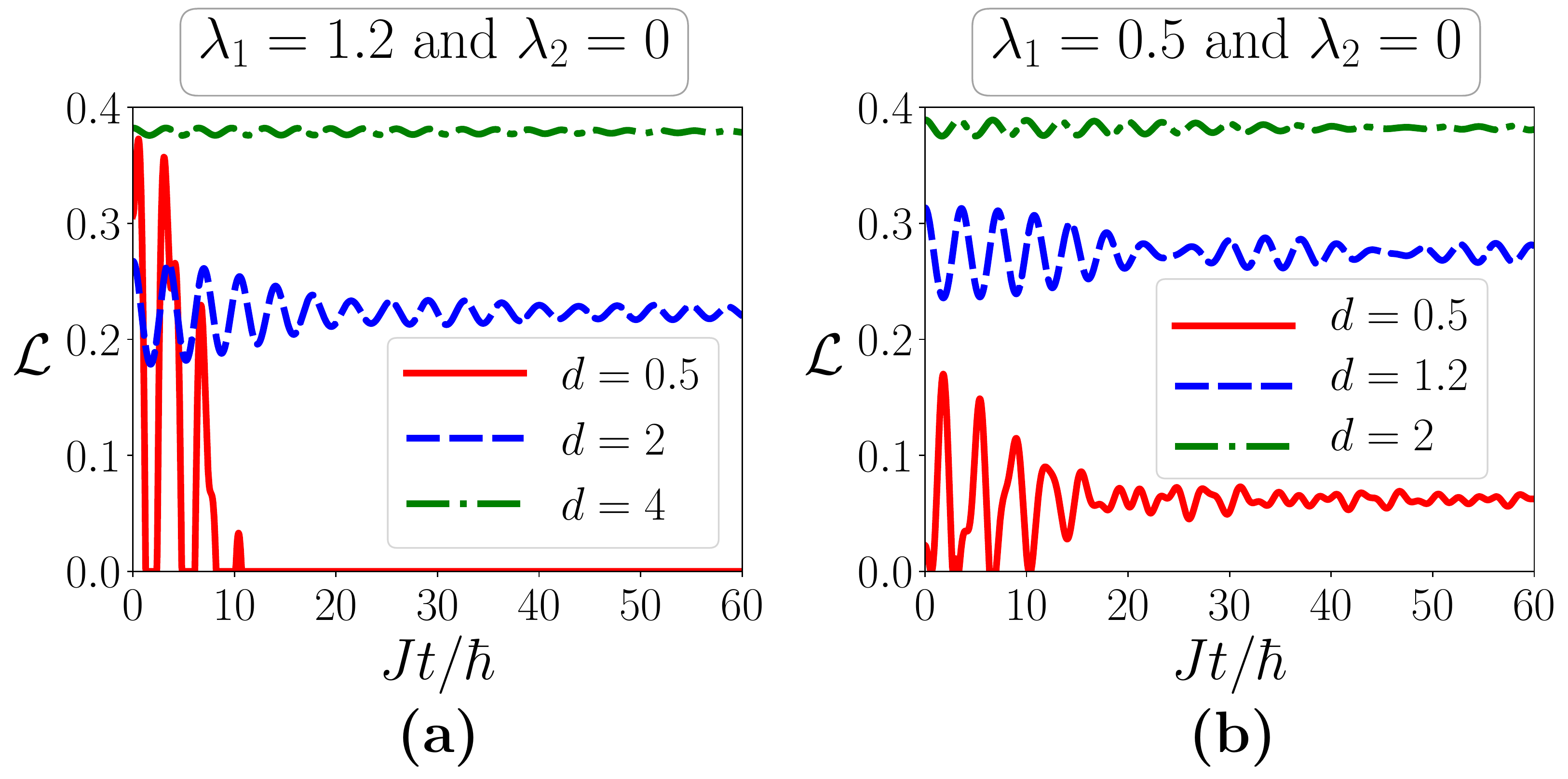}
\caption{(Color online.)  LN against time for different values of $d$, when the initial state is the zero-temperature state. 
The time-evolved state is computed using 
the method discussed in Secs. \ref{sec_model} and \ref{sec_dyn}.
Both the axes are dimensionless.
}
\label{fig:entdynamics}
\end{figure}

\begin{figure*}
\includegraphics[width=\linewidth]{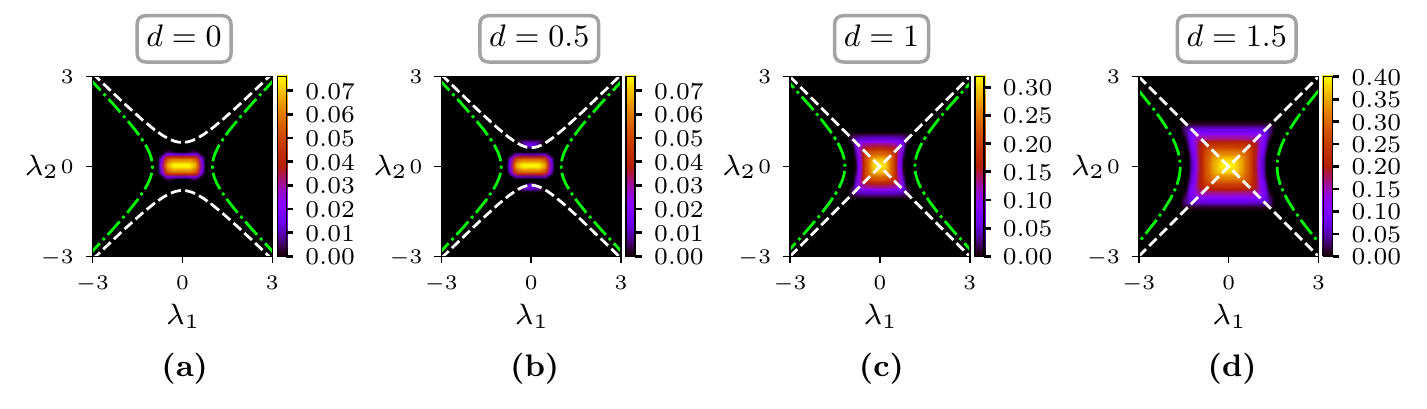}
\caption{(Color online.) Time-averaged entanglement of the time-evolved state (described in Secs. \ref{sec_model} and \ref{sec_dyn}) at large time as function of $\lambda_1$ and $\lambda_2$ for different values of $d$. Averaging is performed from $Jt/\hbar = 80\pi$ to $100\pi$.
Both the axes are dimensionless. }
\label{fig:long_time_ent}
\end{figure*}

First of all, we note that \(m^x(t) = m^y(t) =0\) and \(C^{xz}(t)  = C^{zx}(t) = C^{yz}(t) = C^{zy}(t) =0\) of the evolved state of the DATXY model  for all values of  $t$ and $d$ like in the canonical equilibrium state. We now choose the initial state for the evolution as the zero-temperature state.
Like in the previous static cases, finding a closed form of any physical quantities with time for the DATXY model  is analytically hard. However, if we consider the zero-temperature state of the DUXY model, given in Eq. (\ref{ham_spin}) with \(h_2 =0\), as an initial state,  and  evolve the system according to  \(\hat{H}(t)\), we can  analytically obtain all the two-site classical correlators and transverse magnetizations for $t \geq 0$.
In this case, to obtain $m^z (t)$ and other two-site correlators of the DUXY model with $d\ne 0$, let us first define the quantities $K_R(\tilde{t},\phi_p)$, $S(\tilde{t},\phi_p)$ and $M(\tilde{t},\phi_p)$ as follows.
\footnotesize
\begin{eqnarray}
K_R(\tilde{t},\phi_p) &=&  \frac{\gamma}{\pi} \frac{\sin(\phi_p R)\sin \phi_p}{\Lambda(\lambda_1)\Lambda^2(0)}
\Big[\gamma^2 \sin^2\phi_p+(\cos\phi_p+\lambda_1)\cos\phi_p \nonumber \\ 
&-&\lambda_1\cos\phi_p\cos\big(2\Lambda(0)\tilde{t}\big)\Big] \nonumber\\
&-&\frac{1}{\pi}\frac{\cos\phi_p}{\Lambda(\lambda_1) \Lambda^2(0)} \Big[ \big(\gamma^2\sin^2\phi+ (\cos\phi_p+\lambda_1)\cos\phi_p\big)\cos\phi_p \nonumber \\
&+& \lambda_1\gamma^2\sin^2\phi_p\cos\big(2\Lambda(0)\tilde{t}\big) \Big],
\end{eqnarray}
\small
\begin{eqnarray}
S(\tilde{t},\phi_p) &=&  \frac{\gamma \lambda_1}{\pi}  \sin^2 \phi_p \frac{\sin[2\tilde{t}\Lambda(0)]}{\Lambda(\lambda_1) \Lambda(0)} ,
\end{eqnarray}
and 
\begin{eqnarray}
M(\tilde{t},\phi_p) &=& -\frac{1}{\pi}\frac{1}{\Lambda(\lambda_1) \Lambda^2(0)} 
 \Big[\cos\big(2 \Lambda(0)\tilde{t}\big) \lambda_1 \gamma^2 \sin^2\phi_p \nonumber \\
 &+&  \cos\phi \big(\gamma^2 \sin^2 \phi_p + (\cos\phi_p + \lambda_1)\cos\phi_p \big)\Big].
\end{eqnarray}
\normalsize
Here \(\Lambda(x)= \sqrt{\gamma^2\sin^2\phi_p + (x+\cos\phi_p)^2}\) and
$\tilde{t} = Jt/\hbar$.
In terms of \(K_R(\tilde{t},\phi_p)\), \(S(\tilde{t},\phi_p)\), and \(M(\tilde{t},\phi_p) \), we can express all  the classical correlators and  magnetizations of the time-evolved state (See Table \ref{t2}).
Following Eq. \eqref{rho2},  the nearest-neighbor two-site density matrix and hence bipartite entanglement (LN) of the evolved state can be computed. 
The time-evolved state is again insensitive to $d$ when  the evolution starts with the zero-temperature state and \(d < \gamma\). The picture remains qualitatively similar for any other initial thermal state with moderate values of \(\beta\), which can be obtained through numerical simulations. In case of the DATXY model, we can numerically compute magnetizations and two-site correlators, and thus
bipartite entanglement for all values of $t$ and $d$.

We now discuss the pattern of entanglement with time according to the phases of the initial state with different values of $d$.
\begin{enumerate}
\item When the initial state is in a PM (PM-I or PM-II) phase with $d=0$, entanglement fluctuates for small values of $t$ and then vanishes for large time irrespective of $\gamma$ and $(\lambda_1, \lambda_2)$-pair \cite{atxy_pra}.  If we now switch on the DM interaction, i.e.\ for small values of $d$, the situation remains qualitatively similar. However, for moderate values of $d$, especially when $d$ and $\gamma$ posses comparable values,
 not only entanglement content increases for the entire duration, it saturates to a positive value for a large time after  
some initial fluctuations. 
The converging value of entanglement for $t \rightarrow \infty$ increases with the increase of the strength of $d$. It clearly shows the usefulness of the model with DM interaction from the perspective of quantum information processing tasks. 

\item If the system is initially  in the AFM phase except the regions close to the PM-I or PM-II boundaries,  entanglement persists for a large time even without DM interaction \cite{atxy_pra}. However, if the DM interaction is  stronger than the values of $(\lambda_1, \lambda_2)$-duo or comparable, we find that \(\mathcal{L}^{d \gtrsim (\lambda_1, \lambda_2)}(t) >  \mathcal{L}^{d =0} (t)\) for large $t$, where \(\mathcal{L}^{d \gtrsim (\lambda_1, \lambda_2)} (t)\) and  \( \mathcal{L}^{d =0} (t)\) denote LN with $d$ being stronger than or comparable to $(\lambda_1, \lambda_2)$ pair and that without DM, respectively (see Fig. \ref{fig:entdynamics}). 
\end{enumerate}

To summarize,  with $d=0$, entanglement sustains for large time only in a small region of the AFM phase. With increasing values of $d$, the regions of non-vanishing entanglement at large time increases in the $(\lambda_1, \lambda_2)$-plane (see Fig. \ref{fig:long_time_ent}). Moreover, the value of  entanglement at large time also increases with increasing value of $d$. Therefore, the advantage of introducing DM interaction can  clearly be observed from the production of sustainable entanglement at large time.

 Let us now discuss the statistical properties of physical quantities in this scenario. In a quantum mechanical system, ergodic theorem states that time average of any properties should match with that of the ensemble average \cite{ergo_klein,ergo_loinger, bm1, bm2} Otherwise the physical quantity is said to be non-ergodic. A given physical property, \(\mathcal{P}\), is said to be ergodic if we can find a temperature $T'$, such that its thermal equilibrium value at large time (i.e., $\mathcal{P}_{eq}(T', \lambda_1^{\infty},\lambda_2^{\infty})$) is equal to its time averaged value at large time $\mathcal{P}_{\infty}(T,\lambda_1,\lambda_2)$, where $T$ is the initial temperature of the system. In other words, $\mathcal{P}$ will be ergodic if we have 
 \begin{eqnarray}
\underset{T'}{\mbox{max}} \left[\mathcal{P}_{eq}(T',\lambda_1^{\infty},\lambda_2^{\infty})\right] \geq \mathcal{P}_{\infty}(T,\lambda_1,\lambda_2). 
 \end{eqnarray}
  Otherwise, $\mathcal{P}$ will be non-ergodic.

\noindent\textbf{Theorem 3.}
\emph{Bipartite entanglement in the DATXY model is always ergodic.}

\noindent\textbf{Proof.} 
Bipartite entanglement was shown to be ergodic  as well as nonergodic in case of ATXY model without DM interaction \cite{atxy_pra}. 
As depicted in Fig. \ref{fig:long_time_ent}, we observe that the maximum value of time averaged LN occurs for $\lambda_1 = \lambda_2 = 0$ for all values of $d$. Moreover, at $\lambda_1 = \lambda_2 = 0$ point, the system is not perturbed with time for the quench mentioned in Eq. (\ref{eq:quench}). Therefore, we have $\mathcal{L}_{\infty}(T,\lambda_1,\lambda_2) \leq \mathcal{L}_{\infty}(T,0,0) = \mathcal{L}_{eq}(T,\lambda_1^{\infty},\lambda_2^{\infty}) \leq \underset{T'}{\mbox{max}} \left[\mathcal{L}_{eq}(T',\lambda_1^{\infty},\lambda_2^{\infty})\right]$, which proves the ergodic nature of bipartite entanglement in this model, thereby wiping out the nonergodicity with the help of DM interaction. \hfill $\blacksquare$

If we choose other physical quantity, e.g. quantum discord \cite{discord1, discord2}, which is non-ergodic in case of vanishing DM interaction \cite{atxy_pra}, we find that it also becomes ergodic if one sufficiently increases $d$.

\section{Discussion} 
\label{sec_discuss}

Almost 80 years ago, the importance of an anti-symmetric interaction, known as the Dzyaloshinskii-Moriya (DM) interaction, along with symmetric ones was realized towards explaining certain features observed in some solid state systems. 
In this paper, we  explored the static as well as dynamical properties of both the classical as well as quantum correlation of the quantum XY model with transverse uniform and alternating magnetic fields 
 in the presence of  the DM interaction. While the transverse field XY model is an well studied prototypical model in literature, the study of the same in the presence of uniform and alternating transverse fields (ATXY)  has remain mostly 
unexplored, although it offers a richer phase diagram. 

In this work, we systematically explored the effects of DM interaction on different phases of the 
ATXY model by suitably choosing the relevant order parameters. More precisely, when the DM interaction is weaker than the anisotropy present in the exchange interaction, the model possesses one antiferromagnetic (AFM) and two paramagnetic phases (PM-I and PM-II), similar to the ATXY model without the DM interaction. However, in presence of DM interaction and the alternating magnetic field, the AFM has commensurate and incommensurate order depending on the system parameters. Moreover, in absence of the alternating magnetic field, the ground state of the system is \emph{completely insensitive} towards the DM interaction. The situation changes radically, when the DM interaction becomes stronger than the anisotropy of the exchange interaction: a new \emph{gapless} chiral (CH) phase emerges in place of the AFM phase. In this work, we reported all the critical lines present between these four different quantum phases.

We think that the results presented in this paper can be interesting from a fundamental 
point of view which deals with quantum phase transitions. With the recent progress in current technologies, we are hopeful that in future, such model can be 
prepared in laboratories, in particular, with cold-atomic substrates. For example, with ion-trap, inhomogeneous transverse magnetic field can be prepared if individual ions can be addressed separately with appropriate laser field. Note 
here that only nearest-neighbor exchange interaction is difficult to prepare in ion-trap setup as it always ended up with long range power law ($\frac{1}{r^\alpha}$)-type interaction \cite{realization1,realization2}. However, with 
adequately large power exponent ($\alpha$), the essential physics of the nearest-neighbor model can be captured.  Another potential candidate for realizing our 
Hamiltonian  is ultracold atoms in optical lattice\cite{realization3} and cold gas\cite{coldgas1, coldgas2}. Recently it was shown that models with site dependent magnetic fields like disordered systems can be realized in ultracold 
atoms \cite{realization4,realization5}, thereby opening up possibilities to realize models like the one considered here. 
 The choice of the dimension have been made one 
since the model and its physical quantities can be computed analytically without any approximations which is not the case for higher dimensions. In this 
respect, we would like to mention that there are some findings in a different context that the effect of DM interaction does not change the properties substantially by 
increasing the dimension from one to two \cite{master_thesis}.


In Ref \cite{dm17}, the effects of DM interaction in the behaviors of classical and quantum correlations, including entanglement,
in the quantum XY model with uniform transverse magnetic field (UXY) have been explored. On the other hand, we have showed that
all the critical lines, including gapped-to-gapless phase transitions, found  via different order parameters can also be detected by examining the first derivatives of the bipartite entanglement of the zero-temperature state.
Moreover, we found that vanishing entanglement  surfaces in the parameter spaces of the ATXY model now become a volume in presence of DM interaction at zero-temperature. 

We found that introduction of DM interaction  
can be beneficial to obtain durable bipartite entanglement in the time-evolved states and hence  the model can be a suitable candidate for realizing quantum information processing tasks.  Moreover, comparing the behavior of bipartite entanglements of the canonical equilibrium and the time-evolved states, we concluded that the DM interaction invariably induces the bipartite entanglement of this model to be ergodic in nature. %


\acknowledgments
The authors thank Sauri Bhattacharyya, Samrat Kadge, and Dibya Kanti Mukherjee for fruitful discussions, and
acknowledge computations performed at the cluster computing facility of Harish-Chandra Research Institute, Allahabad, India. 
Numerical results have been obtained using the Quantum Information and Computation library 
(QIClib) (\url{https://titaschanda.github.io/QIClib}). DS is supported by the National Science Centre (Poland) under QuantERA programme No. 2017/25/Z/ST2/03028. 
TC was supported in part by Quantera QTFLAG project 2017/25/Z/ST2/03029 of National Science Centre (Poland).
This research was supported in part by the `INFOSYS scholarship for senior students'.

\appendix

\section{Diagonalization of the Hamiltonian for non-zero alternating field}
\label{ap:diagonalization}

The Hamiltonian $\hat{H}_p$, given in Eq. (\ref{hp}), that acts on the $p^{th}$ subspace of dimension $16$ can be block-diagonalized by a choice of basis
$\{|\psi_i\rangle:1,\cdots,16\}$, given by
\begin{eqnarray}
|\psi_{1}\rangle &=& \hat{a}_p^{\dagger}\hat{b}_{p}^{\dagger}|0\rangle,\nonumber\\
|\psi_{2}\rangle &=& \hat{a}_{-p}^{\dagger} \hat{b}_{-p}^{\dagger}|0\rangle,  
\label{psub_1}
\end{eqnarray}
\begin{eqnarray}
|\psi_3\rangle &=& \hat{a}_p^{\dagger}|0\rangle,\nonumber\\
|\psi_4\rangle &=&  \hat{b}_{p}^{\dagger}|0\rangle,\nonumber\\
|\psi_5\rangle &=& \hat{a}_p^{\dagger} \hat{a}_{-p}^{\dagger} \hat{b}_{p}^{\dagger}|0\rangle,\nonumber\\
|\psi_{6}\rangle &=& \hat{a}_p^{\dagger} \hat{b}_{p}^{\dagger}\hat{b}_{-p}^{\dagger}|0\rangle, 
\label{psub_2}
\end{eqnarray}
\begin{eqnarray}
|\psi_{7}\rangle &=& \hat{a}_{-p}^{\dagger}|0\rangle,\nonumber\\
|\psi_{8}\rangle &=& \hat{b}_{-p}^{\dagger}|0\rangle\nonumber\\
|\psi_{9}\rangle &=& \hat{a}_p^{\dagger} \hat{a}_{-p}^{\dagger} \hat{b}_{-p}^{\dagger}|0\rangle,\nonumber\\
|\psi_{10}\rangle &=& \hat{a}_{-p}^{\dagger} \hat{b}_{p}^{\dagger}\hat{b}_{-p}^{\dagger}|0\rangle,
\label{psub_3}
\end{eqnarray}
\begin{eqnarray}
|\psi_{11}\rangle &=& |0\rangle, \nonumber \\
|\psi_{12}\rangle &=& \hat{a}_p^{\dagger} \hat{b}_{-p}^{\dagger}|0\rangle\nonumber\\
|\psi_{13}\rangle &=& \hat{a}_{-p}^{\dagger} \hat{b}_{p}^{\dagger}|0\rangle\nonumber\\
|\psi_{14}\rangle &=& \hat{a}_p^{\dagger} \hat{a}_{-p}^{\dagger}|0\rangle,\nonumber\\
|\psi_{15}\rangle &=& \hat{b}_p^{\dagger} \hat{b}_{-p}^{\dagger}|0\rangle,\nonumber\\
|\psi_{16}\rangle &=& \hat{a}_p^{\dagger}\hat{a}_{-p}^{\dagger} \hat{b}_{p}^{\dagger}\hat{b}_{-p}^{\dagger}|0\rangle,
\label{psub_4}
\end{eqnarray}
where $|0\rangle$ denotes the vacuum state of the Fermi operators, $\hat{a}^{\dagger}_p$ and $\hat{b}^{\dagger}_p$. 
Note that the above sets of basis block-diagonalize $\hat{H}_p$ into four blocks of dimensions $2$, $4$, $4$, and $6$, such that 
$\hat{H}_p=\bigoplus_{k=1}^{4}\hat{H}_p^k$. Using the form of 
$\hat{H}_p$, given in  Eq. (\ref{hp}), and Eqs. (\ref{psub_1})-(\ref{psub_4}), $\hat{H}_p^1$ is found to be a null matrix of dimension $2$, while
\small
\begin{widetext}
\begin{eqnarray}
\hat{H}_p^2&=& J
\left[
\begin{array}{cccc}
-\lambda_1-\lambda_2 & \cos{\phi_p} + d\sin\phi_p & -i\gamma \sin{\phi_p} & 0 \\
 \cos{\phi_p} + d\sin\phi_p & -\lambda_1+\lambda_2 & 0 & -i\gamma \sin{\phi_p}\\
i\gamma\sin{\phi_p} & 0 & \lambda_1-\lambda_2 & -\cos\phi_p + d\sin\phi_p \\
 0 & i\gamma\sin{\phi_p} & -\cos\phi_p+d\sin\phi_p & \lambda_1+\lambda_2  
\end{array}
\right],
\label{Hp2}
\end{eqnarray}
\begin{eqnarray}
\hat{H}_p^3&=& J
\left[
\begin{array}{cccc}
-\lambda_1-\lambda_2 & \cos{\phi_p} - d\sin\phi_p & -i\gamma \sin{\phi_p} & 0 \\
 \cos{\phi_p} - d\sin\phi_p & -\lambda_1+\lambda_2 & 0 & -i\gamma \sin{\phi_p}\\
i\gamma\sin{\phi_p} & 0 & \lambda_1-\lambda_2 & -\cos\phi_p - d\sin\phi_p \\
 0 & i\gamma\sin{\phi_p} & -\cos\phi_p - d\sin\phi_p & \lambda_1+\lambda_2  
\end{array}
\right],
\label{Hp3}
\end{eqnarray}
\begin{eqnarray}
\hat{H}_p^4&=& J
\left[
\begin{array}{cccccc}
-2\lambda_1 & i\gamma \sin{\phi_p} & -i\gamma \sin{\phi_p} & 0 & 0 &0 \\
 -i\gamma\sin{\phi_p} & 0 & 0 & \cos{\phi_p} - d\sin\phi_p & \cos{\phi_p} + d\sin\phi_p & -i\gamma \sin{\phi_p}\\
i\gamma\sin{\phi_p} & 0 & 0 & -\cos{\phi_p} - d\sin\phi_p & - \cos{\phi_p} + d\sin\phi_p & i\gamma \sin{\phi_p}\\
0 & \cos{\phi_p} - d\sin\phi_p & -\cos{\phi_p} - d\sin\phi_p & -2\lambda_2 & 0 & 0 \\
0 & \cos{\phi_p} + d\sin\phi_p & -\cos{\phi_p} + d\sin\phi_p & 0 & 2\lambda_2 & 0 \\
0 & i\gamma \sin{\phi_p} & -i\gamma \sin{\phi_p} & 0 & 0 & 2\lambda_1
\end{array} 
\right].
\label{Hp4}
\end{eqnarray}
\end{widetext}
\normalsize
Hence, diagonalization of the $p^{th}$ subspace of dimension $16$ reduces to the diagonalization of the 
matrices $\tilde{H}_p^k$, $k=1,2,3,4$. 

\section{Two-site spin correlators and magnetizations for non-zero alternating field}
\label{ap:corr_fun}
Similar to the Hamiltonian $\hat{H}_p$, the two-site spin correlators, $\hat{C}^{\alpha\mu}_p$, with $\alpha, \mu=x,y$, mentioned in Eqs. (\ref{eq:corr_op}) and (\ref{eq:alt_cc}), can be block-diagonalized in the same basis given in Eqs. (\ref{psub_1}) - (\ref{psub_4}).
For example, one can write $\hat{C}^{xx}_p=\bigoplus_{k=1}^4\hat{C}^{xx,k}_p$, 
where $\hat{C}^{xx,1}_p$ is a 
null matrix of dimension $2$, and $\hat{C}^{xx,2}_p$, $\hat{C}^{xx,3}_p$, and $\hat{C}^{xx,4}_p$ are respectively given by  
\begin{widetext}
 \begin{eqnarray}
 \hat{C}^{xx,2}_p&=& 
\left[
\begin{array}{cccc}
0 & e^{i\phi_p} & -e^{i\phi_p} & 0  \\
    e^{-i\phi_p} & 0 & 0 & e^{-i\phi_p}  \\
    -e^{-i\phi_p} & 0 & 0 & -e^{-i\phi_p} \\
    0 & e^{i\phi_p} & -e^{i\phi_p} & 0 
\end{array}
\right],\,\,
\hat{C}^{xx,3}_p=  
\left[
\begin{array}{cccc}
0 & e^{-i\phi_p} & e^{-i\phi_p} & 0  \\
    e^{i\phi_p} & 0 & 0 & -e^{i\phi_p}  \\
    e^{i\phi_p} & 0 & 0 & -e^{i\phi_p} \\
    0 & -e^{-i\phi_p} & -e^{-i\phi_p} & 0 
\end{array}
\right],\nonumber\\
  \hat{C}^{xx,4}_p&=&
  \left[ 
  \begin{array}{cccccc}
    0 & -e^{-i\phi_p} & -e^{i\phi_p} & 0 & 0 & 0 \\
    -e^{i\phi_p} & 0 & 0 & e^{i\phi_p} & e^{i\phi_p} & -e^{i\phi_p} \\
    -e^{-i\phi_p} & 0 & 0 & -e^{-i\phi_p} & -e^{-i\phi_p} & -e^{-i\phi_p} \\
    0 & e^{-i\phi_p} & -e^{i\phi_p} & 0 & 0 & 0 \\
    0 & e^{-i\phi_p} & -e^{i\phi_p} & 0 & 0 & 0 \\
    0 & -e^{-i\phi_p} & -e^{i\phi_p} & 0 & 0 & 0 \\
    \end{array}
   \right].  
 \end{eqnarray}
\end{widetext}
Similar calculation for $\hat{C}^{yy}$ leads to $\hat{C}^{yy,1}_p=\hat{C}^{xx,1}_p$, and 
\begin{widetext}
\begin{eqnarray}
    \hat{C}^{yy,2} &=&
    \left[ 
    \begin{array}{cccc}
    0 & e^{i\phi_p} & e^{i\phi_p} & 0  \\
    e^{-i\phi_p} & 0 & 0 & -e^{-i\phi_p}  \\
    e^{-i\phi_p} & 0 & 0 & -e^{-i\phi_p} \\
    0 & -e^{i\phi_p} & -e^{i\phi_p} & 0 
    \end{array}
    \right],\,\,
    \hat{C}^{yy,3}_p = 
    \left[
    \begin{array}{cccc}
    0 & e^{-i\phi_p} & -e^{-i\phi_p} & 0  \\
    e^{i\phi_p} & 0 & 0 & e^{i\phi_p}  \\
    -e^{i\phi_p} & 0 & 0 & -e^{i\phi_p} \\
    0 & e^{-i\phi_p} & -e^{-i\phi_p} & 0 
    \end{array}
    \right],\nonumber\\
    \hat{C}^{yy,4}_p &=& 
    \left[
    \begin{array}{cccccc}
    0 & e^{-i\phi_p} & e^{i\phi_p} & 0 & 0 & 0 \\
    e^{i\phi_p} & 0 & 0 & e^{i\phi_p} & e^{i\phi_p} & e^{i\phi_p} \\
    e^{-i\phi_p} & 0 & 0 & -e^{-i\phi_p} & -e^{-i\phi_p} & e^{-i\phi_p} \\
    0 & e^{-i\phi_p} & -e^{i\phi_p} & 0 & 0 & 0 \\
    0 & e^{-i\phi_p} & -e^{i\phi_p} & 0 & 0 & 0 \\
    0 & e^{-i\phi_p} & e^{i\phi_p} & 0 & 0 & 0 \\
    \end{array}
    \right].
\end{eqnarray}
\end{widetext}
And the operators $\hat{C}^{xy}_p$ and $\hat{C}^{yx}_p$ are given by
\begin{widetext}
\begin{eqnarray}
    \hat{C}^{xy,2}_p &=&-i
    \left[ 
    \begin{array}{cccc}
    0 & e^{-i\phi_p} & -e^{-i\phi_p} & 0  \\
    -e^{i\phi_p} & 0 & 0 & e^{i\phi_p}  \\
    e^{i\phi_p} & 0 & 0 & -e^{i\phi_p} \\
    0 & -e^{-i\phi_p} & e^{-i\phi_p} & 0 
    \end{array}
    \right],\,\,
    \hat{C}^{xy,3}_p = -i
    \left[
    \begin{array}{cccc}
    0 & e^{i\phi_p} & e^{i\phi_p} & 0  \\
    -e^{-i\phi_p} & 0 & 0 & -e^{-i\phi_p}  \\
    -e^{-i\phi_p} & 0 & 0 & -e^{-i\phi_p} \\
    0 & e^{i\phi_p} & e^{i\phi_p} & 0 
    \end{array}
    \right],\nonumber\\
    \hat{C}^{xy,4}_p &=& -i
    \left[
    \begin{array}{cccccc}
     0 & -e^{i\phi_p} & -e^{-i\phi_p} & 0 & 0 & 0 \\
    e^{-i\phi_p} & 0 & 0 & e^{-i\phi_p} & -e^{-i\phi_p} & -e^{-i\phi_p} \\
    e^{i\phi_p} & 0 & 0 & -e^{i\phi_p} & e^{i\phi_p} & -e^{i\phi_p} \\
    0 & -e^{i\phi_p} & e^{-i\phi_p} & 0 & 0 & 0 \\
    0 & e^{i\phi_p} & -e^{-i\phi_p} & 0 & 0 & 0 \\
    0 & e^{i\phi_p} & e^{-i\phi_p} & 0 & 0 & 0 \\
    \end{array}
    \right]
\end{eqnarray}
and
\begin{eqnarray}
    \hat{C}^{yx,2}_p &=&-i
    \left[ 
    \begin{array}{cccc}
    0 & -e^{-i\phi_p} & -e^{-i\phi_p} & 0  \\
    e^{i\phi_p} & 0 & 0 & e^{i\phi_p}  \\
    e^{i\phi_p} & 0 & 0 & e^{i\phi_p} \\
    0 & -e^{-i\phi_p} & -e^{-i\phi_p} & 0 
    \end{array}
    \right],\,\,
    \hat{C}^{yx,3}_p = -i
    \left[
    \begin{array}{cccc}
    0 & -e^{i\phi_p} & e^{i\phi_p} & 0  \\
    e^{-i\phi_p} & 0 & 0 & -e^{-i\phi_p}  \\
    -e^{-i\phi_p} & 0 & 0 & e^{-i\phi_p} \\
    0 & e^{i\phi_p} & -e^{i\phi_p} & 0 
    \end{array}
    \right],\nonumber\\
    \hat{C}^{yx,4}_p &=& -i
    \left[
    \begin{array}{cccccc}
    0 & -e^{i\phi_p} & -e^{-i\phi_p} & 0 & 0 & 0 \\
    e^{-i\phi_p} & 0 & 0 & -e^{-i\phi_p} & e^{-i\phi_p} & -e^{-i\phi_p} \\
    e^{i\phi_p} & 0 & 0 & e^{i\phi_p} & -e^{i\phi_p} & -e^{i\phi_p} \\
    0 & e^{i\phi_p} & -e^{-i\phi_p} & 0 & 0 & 0 \\
    0 & -e^{i\phi_p} & e^{-i\phi_p} & 0 & 0 & 0 \\
    0 & e^{i\phi_p} & e^{-i\phi_p} & 0 & 0 & 0 \\
    \end{array}
    \right],
\end{eqnarray}
\end{widetext}
with $\hat{C}^{xy,1}_p$ and $\hat{C}^{yx,1}_p$ being $2\times2$ null matrices. Furthermore, for $\hat{m}^z_{e,p}$ and $\hat{m}^z_{o,p}$, we get
\begin{eqnarray}
\hat{m}^{z,2}_{e,p} &=& \hat{m}^{z, 3}_{e,p} = \mbox{Diag}\big(-2,0,0,2\big), \nonumber \\
\hat{m}^{z,4}_{e,p} &=& \mbox{Diag}\big(-2,0,0,-2,2,2\big),
\end{eqnarray}
\begin{eqnarray}
\hat{m}^{z,2}_{o,p} &=& \hat{m}^{z, 3}_{o,p} = \mbox{Diag}\big(0,-2,2,0\big), \nonumber \\
\hat{m}^{z,4}_{o,p} &=& \mbox{Diag}\big(-2,0,0,2,-2,2\big),
\end{eqnarray}
with $\hat{m}^{z,1}_{e,p}$ and $\hat{m}^{z,1}_{o,p}$ being null matrices of dimension 2.

Note that we can obtain the two-site correlators and magnetizations, in the thermodynamic limit ($N \rightarrow \infty$), by replacing $\frac{2}{N}\sum_{p=1}^{N/4} \rightarrow \frac{1}{\pi}\int_{0}^{\pi/2}$.
For computing the classical correlators and magnetizations in the thermodynamic limit, we  employ doubly adaptive quadrature method\cite{} to perform all the numerical integrations throughout the paper, where we have set a  tolerance to be $\sim 10^{-10}$ to guarantee the convergence.

\section{Two-site spin correlators and magnetizations in the uniform field case}
\label{ap:corr_uni}
In the uniform field case, we use the same procedure to obtain the two-site spin correlators and the magnetization. 
To compute the classical correlators and the magnetization, we define the following operators. 
\begin{eqnarray}
\hat{C}^{\alpha\mu} = \frac{1}{N}\sum_{i=1}^{N}\hat{\sigma}_i^{\alpha}\hat{\sigma}_{i+1}^{\mu},  \
\hat{m}^{z} = \frac{1}{N}\sum_{i=1}^{N}\hat{\sigma}_i^{z},
\label{eq:dm3}
\end{eqnarray}
where $\alpha,\mu=x,y$.  After successive applications of Jordan-Wigner and Fourier transformations, we get $C^{\alpha\mu} = \frac{1}{N} \sum_{p=1}^{N/2}C^{\alpha\mu}_p$, where
\small
\begin{equation}
C^{\alpha\mu}_p = \frac{1}{Z_p}\text{Tr}[\hat{C}^{\alpha\mu}_p \exp(-\beta \hat{H}_p)], \ 
\label{cc1}
\end{equation} 
\normalsize
Here, $\beta = 1/ k_B T$, $T$ being the temperature of the system.
As before, the correlator, $C^{zz}$, can be computed using the Wick's theorem as
\begin{eqnarray}
C^{zz} = (m^z)^2 - C^{xx}C^{yy} + C^{xy}C^{yx}.
\end{eqnarray}
In the basis $\lbrace|0\rangle, \hat{c}_p^\dagger \hat{c}_{-p}^\dagger |0\rangle, \hat{c}_p^\dagger  |0\rangle, \hat{c}_{-p}^\dagger |0\rangle\ \rbrace$,  the correlator operators, $\hat{C}^{\alpha\mu}_p$, with $\alpha, \mu = x, y$, given in Eqs. (\ref{eq:dm3}) and (\ref{cc1}), have the following forms.

\begin{widetext}
\begin{eqnarray}
\hat{C}^{xx}_p &=& 2\begin{bmatrix}
    0 & i \sin\phi_p & 0 & 0  \\
    - i\sin\phi_p & 2\cos\phi_p & 0 & 0 \\
    0 & 0 & \cos\phi_p  & 0 \\
    0 & 0 & 0 & \cos\phi_p 
  \end{bmatrix}, \ \ \
\hat{C}^{yy}_p = 2\begin{bmatrix}
    0 & -i \sin\phi_p & 0 & 0  \\
     i\sin\phi_p & 2\cos\phi_p & 0 & 0 \\
    0 & 0 & \cos\phi_p  & 0 \\
    0 & 0 & 0 & \cos\phi_p 
  \end{bmatrix}, \nonumber \\
\hat{C}^{xy}_p &=& 2\begin{bmatrix}
    0 & -  \sin\phi_p & 0 & 0  \\
     -\sin\phi_p & 0 & 0 & 0 \\
    0 & 0 & \sin\phi_p  & 0 \\
    0 & 0 & 0 & -\sin\phi_p 
  \end{bmatrix}, \ \
\hat{C}^{yx}_p = 2\begin{bmatrix}
    0 & -  \sin\phi_p & 0 & 0  \\
     -\sin\phi_p & 0 & 0 & 0 \\
    0 & 0 & -\sin\phi_p  & 0 \\
    0 & 0 & 0 & \sin\phi_p 
  \end{bmatrix}.
\end{eqnarray}
\end{widetext}
Similarly, we get the magnetization operator, $\hat{m}^z_p$, in the same basis as
\begin{eqnarray}
\hat{m}^z_p = \mbox{Diag}\big(-2,2,0,0\big).
\end{eqnarray}
Note that the two-site correlators and the magnetization, in the thermodynamic limit ($N \rightarrow \infty$) can be obtained in this case by replacing $\frac{1}{N}\sum_{p=1}^{N/2} \rightarrow \frac{1}{2 \pi}\int_{0}^{\pi}$. The closed analytical forms of the correlators and the magnetization for the zero-temperature state (i.e., $\beta \rightarrow \infty$) are given in Table \ref{t11}. For thermal equilibrium state, the forms of the  correlators and the magnetization are as follows:
\begin{widetext}
\small
\begin{eqnarray}
C^{xx} &=& \frac{1}{\pi}\int_{0}^\pi d\phi_p  \frac{ \frac{1}{\Lambda_{p}}\Big( - \big(\gamma \sin^2\phi_p + (\cos\phi_p + \lambda1)\cos\phi_p \big)\sinh(\beta J\Lambda_{p})  +\Lambda_{p}  \cos\phi_p \cosh(\beta J \Lambda_{\phi_p}) \Big) 
+\cos\phi_p\cosh(\beta J d\sin\phi_p)}{\cosh(\beta J \Lambda_p) + \cosh(\beta J  d\sin\phi_p)}, \nonumber \\ \nonumber \\
C^{yy} &=& \frac{1}{\pi}\int_{0}^\pi d\phi_p  \frac{\frac{1}{\Lambda_p} \Big(\big(\gamma \sin^2\phi_p - ( \cos\phi_p + \lambda_1)\cos\phi_p \big)\sinh(\beta J \Lambda_p)  +\Lambda_p  \cos\phi_p \cosh(\beta J \Lambda_p) \Big) 
+\cos\phi_p\cosh(\beta J d\sin\phi_p)}{\cosh(\beta J \Lambda_p) + \cosh(\beta J d\sin\phi_p)},  \nonumber \\ \nonumber \\
C^{xy} &=& C^{yx} = \frac{1}{\pi}\int_{0}^\pi d\phi_p\frac{ - \sin\phi_p\sinh(\beta J d\sin\phi_p)}{\cosh(\beta J \Lambda_p) + \cosh(\beta J d\sin\phi_p)}, \ \ \ \ \
 m^{z} = \frac{1}{\pi}\int_{0}^\pi d\phi_p \frac{ -(\lambda_1 + \cos\phi_p)\sinh(\beta J \Lambda_p)}{\Lambda_p\big(\cosh(\beta J \Lambda_p) + \cosh(\beta J d\sin\phi_p)\big)},
\end{eqnarray}
\end{widetext}
where $\Lambda_p$ is given in Table \ref{t11}.

%
%

\bibliography{bib}

\begin{thebibliography}{170}
\bibitem [{c_c()}]{c_cyclic}%
  \BibitemOpen
  \bibinfo {note} {We impose ``c-cyclic" boundary condition \cite{lsm}. In the
  thermodynamic limit, the changes due to c-cyclic boundary condition are not
  significant.}\BibitemShut {Stop}%
\bibitem [{phi()}]{phi_def}%
  \BibitemOpen
  \bibinfo {note} {$\phi_1$ and $\phi_2$ are the solution of the equation
  \begin{eqnarray*} (\cos\phi_p + \lambda_1)^2 + \gamma^2 \sin^2\phi_p = d^2
  \sin\phi_p, \end{eqnarray*} such that $0 \leq \phi_1 < \phi_2 \leq \pi$. For
  different system parameters, $\phi_1$ and $\phi_2$ are found to be the
  following. \begin{eqnarray} \phi_1 &=& \tilde{\phi}_1 \text{ and } \phi_2 =
  \tilde{\phi}_2, \text{ when } -1\leq\lambda_1\leq1, \nonumber \\ \phi_1 &=&
  \tilde{\phi}_1 \text{ and } \phi_2 = 2\pi -\tilde{\phi}_2, \text{ when }
  \lambda_1 > 1, \nonumber \\ \phi_1 &=& -\tilde{\phi}_1 \text{ and } \phi_2 =
  \tilde{\phi}_2, \text{ when } \lambda_1<-1, \nonumber \end{eqnarray} where
  \begin{eqnarray} \tilde{\phi}_1 = \cos^{-1}(\frac{-\lambda_1}{\sqrt{1 + d^2 -
  \gamma^2}}) - \cos^{-1}(\frac{1}{\sqrt{1 + d^2 - \gamma^2}}) \nonumber \\
  \tilde{\phi}_2 = \cos^{-1}(\frac{-\lambda_1}{\sqrt{1 + d^2 - \gamma^2}}) +
  \cos^{-1}(\frac{1}{\sqrt{1 + d^2 - \gamma^2}}). \nonumber \end{eqnarray} Note
  that real solutions of $\phi_{1}$ and $\phi_{2}$ exist only when $d > \gamma$
  and $\lambda_1^2 \leq 1 + d^2 - \gamma^2$.}\BibitemShut {Stop}%
\bibitem [{ph_()}]{ph_sym}%
  \BibitemOpen
  \bibinfo {note} {This is true, since the Hamiltonian respects particle-hole
  symmetry.}\BibitemShut {Stop}%
\bibitem [{dmt()}]{dmtopmii}%
  \BibitemOpen
  \bibinfo {note} {Note that in the earlier papers \cite{atxy_pra, atxy_1,
  atxy_2, atxy_3, qptbook3}, the PM-II phase was mentioned as the dimer phase.
  Here, our analysis shows that the dimer order parameter \cite{dm_order1,
  dm_order2} in this phase vanishes and it is indeed paramagnetic in nature.
  However, such finding does not affect the other results in \cite{atxy_pra,
  atxy_1, atxy_2, atxy_3, qptbook3}.}\BibitemShut {Stop}%
\end{thebibliography}

\end{document}